\documentclass[
  reprint,
  amsmath,amssymb,
  aip,
  jasa
]{revtex4-2}
\usepackage{braket}
\usepackage{graphicx}
\usepackage{chemfig}
\usepackage{amsmath}
\usepackage{amsfonts}
\usepackage{physics}
\usepackage{chngcntr}
\usepackage{float}
\usepackage{comment}
\usepackage[normalem]{ulem}
\usepackage{soul}
\usepackage{hyperref}
\usepackage{comment}
\usepackage{array}

\DeclareMathOperator{\Ai}{Ai}
\usepackage{svg}
\usepackage{float}

\DeclareSymbolFont{matha}{OML}{txmi}{m}{it}
\DeclareMathSymbol{\varv}{\mathord}{matha}{118}

\definecolor{OliveGreen}{rgb}{0,0.6,0}

\begin{document}
\title{Anisotropic Cylindrical Waves in a Square Lattice of Acoustic Waveguides
}

\author{I.~Ioannou Sougleridis}\thanks{Now at: PMMH, ESPCI Paris, Université PSL, CNRS, Sorbonne Université, Université Paris Cité, Paris, France}
\affiliation{Laboratoire d’Acoustique de l’Universit\'{e} du Mans (LAUM), UMR 6613, Institut d'Acoustique - Graduate School (IA-GS), CNRS,  Le Mans Universit\'{e}, France}
\affiliation{Department of Physics, National and Kapodistrian University of Athens, University Campus, GR-157 84 Athens, Greece}
\author{O.~Richoux}  
\affiliation{Laboratoire d’Acoustique de l’Universit\'{e} du Mans (LAUM), UMR 6613, Institut d'Acoustique - Graduate School (IA-GS), CNRS,  Le Mans Universit\'{e}, France}
\author{V.~Achilleos}  
\affiliation{Laboratoire d’Acoustique de l’Universit\'{e} du Mans (LAUM), UMR 6613, Institut d'Acoustique - Graduate School (IA-GS), CNRS,  Le Mans Universit\'{e}, France}
\author{G.~Theocharis}  
\affiliation{Laboratoire d’Acoustique de l’Universit\'{e} du Mans (LAUM), UMR 6613, Institut d'Acoustique - Graduate School (IA-GS), CNRS,  Le Mans Universit\'{e}, France}
\author{D. J.~Frantzeskakis}  
\affiliation{Department of Physics, National and Kapodistrian University of Athens, University Campus, GR-157 84 Athens, Greece}

\begin{abstract}
We investigate the propagation of cylindrical waves in a square network of acoustic waveguides. We establish, both theoretically and experimentally, the anisotropic dispersion relation governing wave propagation in the network, and demonstrate excellent agreement between experimental measurements and theoretical predictions. Owing to this anisotropic band structure, each propagation direction exhibits distinct dispersive properties. Consequently, the network supports anisotropic cylindrical waves at both low- and high-amplitudes, with waveforms that vary markedly with direction: from nearly dispersionless pulses to Airy-like wave packets in the linear regime, and from sharp shock-like fronts to smooth solitary-like profiles in the nonlinear regime. The theoretical results are further corroborated by numerical simulations based on the two-dimensional Westervelt equation.
%
%
\end{abstract}
\maketitle

\section{Introduction}

Over the past years, structured materials have been widely studied and developed to manipulate wave motion. In that respect, acoustic metamaterials, namely structured materials made of resonant building blocks, play an important role in the design of various classical wave systems. Earlier studies on acoustic metamaterials based on acoustic waveguides incorporating resonant elements (e.g., Helmholtz resonators \cite{sugimoto_dispersion,sugimoto_dispersion_2} or quarter-wavelength resonators \cite{bradley1,bradley2,bradley3}) paved the way for  a variety of important applications. These include  acoustic diodes \cite{diode}, perfect absorbers \cite{vassos_nonlinear,perfect_absorption_1,metasurfaces_1}, acoustic lenses for sub-diffraction imaging \cite{lens}, acoustic sound focusing based on gradient index lenses \cite{gradient,gradient_2}, acoustic topological systems \cite{Ma,coutant2021acoustic} acoustic cloaking \cite{cloaking_1,cloaking_2,cloaking_3,cloaking_4}, bifurcation-based acoustic switching or
rectification \cite{bifurcation}, and so on. 

The above studies predominantly focus on one-dimensional (1D) settings and linear wave phenomena. Nevertheless, nonlinear wave dynamics in 1D airborne acoustic metamaterials have been explored in a growing number of works. These studies were primarily motivated by the seminal contributions of Sugimoto and co-workers, who demonstrated the formation of acoustic solitons in air-filled waveguides side-loaded with Helmholtz resonators \cite{sugimoto_prl,sugimoto_review} (see also \cite{wave_motion,vassos_soliton}). In such systems, dispersion arises from the local resonances of the Helmholtz resonators, as well as Bragg scattering due to periodicity, while nonlinearity originates from finite-amplitude sound propagation in air. In the long-wavelength and small-amplitude limits, the resulting dynamics are governed by an effective Korteweg–de Vries (KdV) equation \cite{ablowitz}. Beyond the pulse-like KdV-type solitary waves, it was also shown that the waveguide side-loaded with Helmholtz resonators supports envelope solitons, which were described by an effective nonlinear Schrödinger (NLS)  \cite{vassos_soliton}. Similar envelope soliton solutions have been predicted in other 
1D acoustic metamaterial configurations, including waveguides loaded with elastic membranes or side holes, where dispersion --induced by periodicity and local resonances-- is combined with weak nonlinearity to yield effective nonlinear Klein–Gordon and NLS-type 
equations \cite{kinezoula_dark,kinezoula_gap}.

In contrast, nonlinear effects in mechanical and elastic metamaterials have been extensively investigated in both one- and two-dimensional (2D) settings \cite{elastic_metamaterials,deng2021nonlinear,tournat2024,vector_mechanical,demiquel,DEMIQUEL2024103394,PALIOVAIOS2024102199}, whereas airborne acoustic metamaterials remain far less explored in two dimensions. Extending nonlinear airborne acoustic metamaterials beyond 1D is expected to lead to substantially richer wave dynamics due to multidirectional propagation, lattice geometry, and symmetry effects. For instance, very recently \cite{ioannou2025ring}, linear and nonlinear waves with radial symmetry, including ring-shaped solitons, were predicted to occur  
%
in a 2D acoustic network consisting of a square lattice of waveguides loaded with Helmholtz resonators at the junctions; 
the resonators were shown to suppress the inherent anisotropy of the square lattice (see also Refs.~\cite{depollier,acousticgraphene,nikitenkova2022symmetric} for related studies).

Inspired by the above developments, in the present work we investigate nonlinear wave phenomena in 
a square network of acoustic waveguides, in the absence of the Helmholtz resonators. We show that the considered setting features an underlying band structure with an  intrinsic anisotropy, 
similarly to other setups in water waves \cite{maurel2017revisiting,pham2025homogenized}, elasticity \cite{auffray2015complete,wautier,burz_2d_linear,burz_cylindrical} and electric transmission lines \cite{stepanyants1981,nikitenkova2022symmetric}. 
%
To study the interplay between anisotropic  dispersion and nonlinearity, 
we extend the analytical framework based on the 2D 
gas-dynamics equations and the electroacoustic analogy by introducing a refined discretization scheme, which results in an 
effective partial differential equation (PDE) valid in the long-wavelength regime, namely a 2D Boussinesq equation. This model features an explicit angle-dependent dispersive coefficients, reflecting the directional variation of the lattice dispersion relation, and is used to   
describe weakly nonlinear and weakly dispersive waves in the square network. 

%

Our analysis reveals a rich family of anisotropic cylindrical waves, whose characteristics vary continuously with the propagation direction. In the linear limit, the model yields angle-dependent dispersive cylindrical waveforms that generalize the self-similar Airy-type solutions found in \cite{ioannou2025ring}, and recover the standard far-field behavior of the 
2D wave equation along the lattice diagonals. In the nonlinear regime, the same framework predicts anisotropic cylindrical pulses ranging from smooth solitary
waves \cite{ioannou2025ring,stepanyants1981,zhang2024solitons} to shock-like structures,
depending on the propagation direction. These predictions are validated by direct numerical simulations of the full lattice dynamics, demonstrating the coexistence of distinct nonlinear propagation regimes along different directions. Overall, 
our results reveal how the anisotropy profoundly influences 
the evolution of radially symmetric linear and nonlinear 
waves, and establish a general
framework for predicting direction-dependent wave phenomena in 
2D acoustic metamaterials.

%
%


A brief presentation of our methodology and findings, along with the description of the organization of the manuscript, 
are as follows. In Sec.~II, we present 
the square acoustic network under consideration, as well as our experimental setting. We experimentally obtain the dispersion relation of the square network, which is then verified analytically and numerically. The results show that, in the long-wavelength regime, the dispersion of the  network is inherently anisotropic, while the propagation along the angle $\theta=\pi/4$ is dispersionless. In addition, we introduce an improved electroacoustic analogy (EA) through the fluid conservation laws, and a fine discretization of the square network. This approach leads to an effective 2D Boussinesq equation, which is shown to accurately capture the anisotropic dispersive behavior of the network, even for shorter-wavelengths. Next, it is shown (also in Sec.~III), that 2D effective Boussinesq in the limit of large-radii and small-amplitude (weakly nonlinear) waveforms, is then reduced to the cylindrical Korteweg-de Vries (cKdV) equation, with an angle-dependent dispersive coefficient. Furthermore, in Sec.~IV, we present linear and nonlinear, angle dependent, cylindrical solutions of the effective anisotropic cKdV. The former, in the linear regime, for $\theta\neq0$ are self-similar solutions of the linearized cKdV exhibiting an Airy-function profile, while for $\theta\to0$ we obtain the free space (dispersionless) solution. The latter, in the nonlinear regime, for $\theta\neq0$, are cylindrical solitons, while for $\theta\to0$, the cKdV is reduce to the inviscid radial Burgers equation, which supports the formation of cylindrical shock waves . In the same Section (Sec.~IV) we present results of direct numerical simulations, for both linear and nonlinear anisotropic cylindrical solutions in the acoustic network. 
Finally, in Sec.~V, we summarize our findings and propose future research directions.

\section{Acoustic square network}

We consider a square lattice composed of simply connected acoustic waveguides of cross-sectional area $S_w$, arranged in a square pattern with lattice spacing $d$, as 
illustrated in Fig.~\ref{square sketch}(a); 
the unit cell is depicted in panel (b).
%
%
Note that considerable attention has been devoted to similar 2D networks \cite{depollier,acousticgraphene,zheng2019,coutant_robustness}, as they provide an excellent platform for manipulating wave propagation in two dimensions and offer high adaptability through variations in symmetry, geometry, and incorporation of local resonances. 
%

\subsection{
Dispersion relation of the square network}
\begin{figure}[tbp!]
\begin{center}
\includegraphics[width=0.5\textwidth]{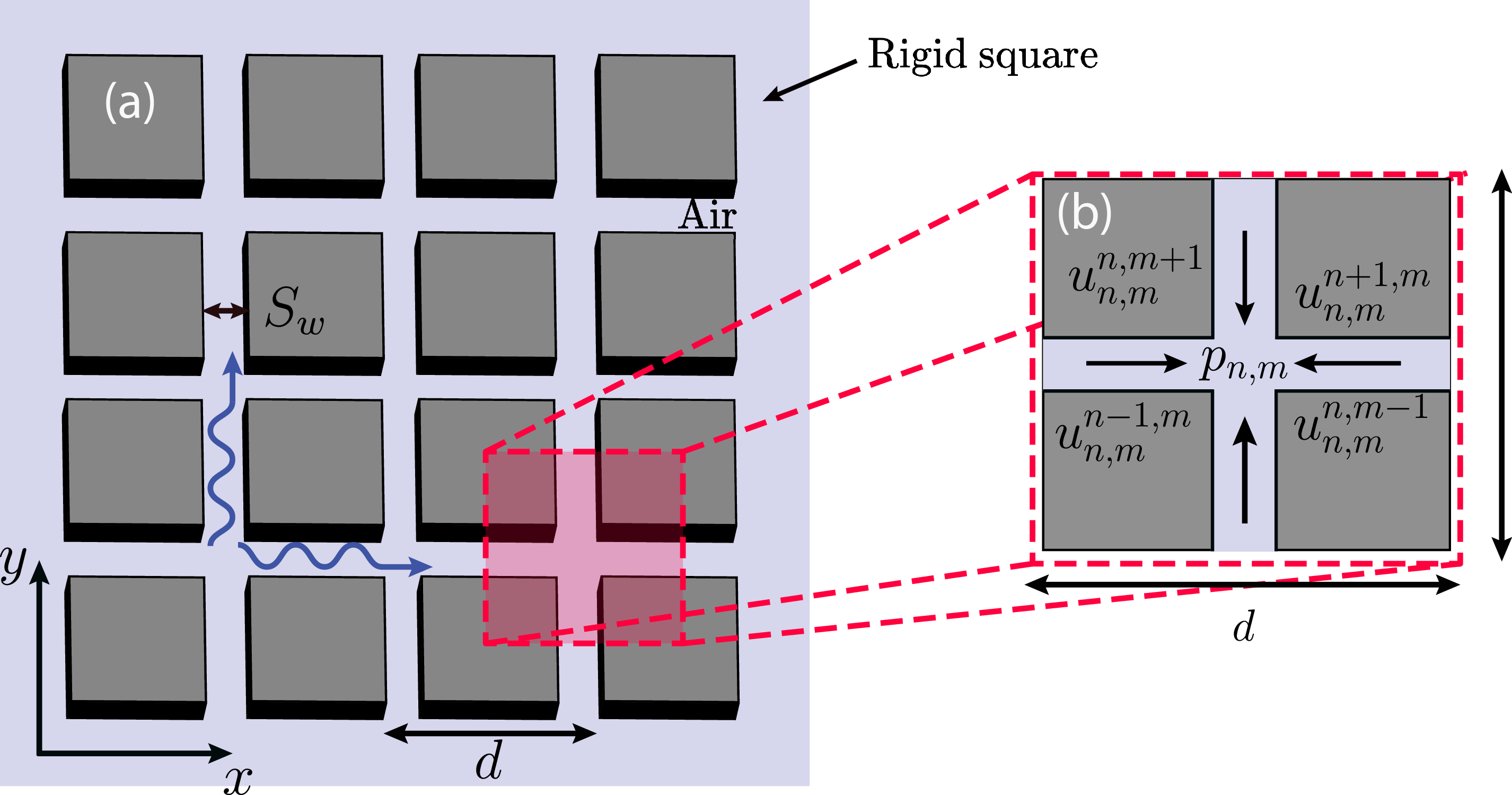} 
 \caption{Schematic of the acoustic periodic network. (a) A periodic 2D square network, composed by narrow waveguide channels of equal length $d$ and cross-section $S_w$, depicted by dark grey color. Depicted by a red rectangle is the unit cell of the network. (b) A zoom of the unit cell with length $d$. The discretized pressure field in the center and the black arrows denote the ingoing flows.  \label{square sketch}}
\end{center}
\end{figure}
Following 
considerations for an ideal fluid in the linear regime, we neglect nonlinearity, viscosity and other dissipative effects. 
Hence, 
the acoustic pressure field $p(x,y,z)$ inside the acoustic network, illustrated in Fig.\,\ref{square sketch}(a), is governed by the three-dimensional (3D)  Helmholtz equation with Neumann boundary conditions, corresponding to zero normal velocity at the rigid walls $\partial_n p=0$. 

To determine the dispersion relation of the square network we work as follows. We consider 
the long-wavelength limit --where the wavelength is much larger than the waveguide cross section-- 
and assume that only the plane mode propagates within each waveguide. Consequently, wave propagation between network junctions can be described (in the linear regime) by the 1D Helmholtz equation \cite{depollier,topology_2D,acousticgraphene}.
Hence, under the monomodal approximation, 
conservation of flux in the central junction of the unit cell reads 
\begin{equation}
u_{n, m}^{n-1, m}+u_{n, m}^{n, m+1}+u_{n, m}^{n, m-1}+u_{n, m}^{n+1, m}=0, \label{flux conservation}
\end{equation}
where $u_{n, m}^{n\pm1, m\pm1}$ are the fluxes incoming from the nodes $(n\pm1, m\pm1)$ to the node $(n,m)$.
Next, we employ the transfer matrix method (TMM) to express the pressure at the junction $(n,m)$, $p_{n,m}$,  and the incoming flux $u^{l,j}_{n,m}$ as a function of the pressure at the junction $(l,j)$, $p_{l,j}$, and the corresponding flux $u^{n,m}_{l,j}$
\begin{align}
    \left[\begin{array}{c}
p_{n,m} \\
u^{i,j}_{n,m}
\end{array}\right]=\mathbf{M}\left[\begin{array}{l}
p_{i,j} \\
-u^{n,m}_{i,j}
\end{array}\right]. \label{transfer definition square}
\end{align}
Here, $\mathbf{M}$ is the transfer matrix for a uniform waveguide segment of length $d$, located between 
the junction $(i,j)$ 
and the junction $(n,m)$. For the square lattice under consideration, 
indices $(l,j)$ are $l=n\pm1$ and $j=m\pm1$.
After some algebra, we can write the incoming fluxes $u_{n, m}^{l, j}$ as
\begin{equation}
u_{n, m}^{l, j}=\displaystyle{\mathrm{i}\left[\displaystyle{\frac{1}{\sin{\left(\omega\frac{d}{c_0}\right)}}} p_{l, j}-\tan{\left(\omega\frac{d}{c_0}\right)} p_{n, m}\right].} \label{junctions}
\end{equation} 
Substituting Eq.~\eqref{junctions} 
for the pressure of the incoming fluxes  
into Eq.\,\eqref{flux conservation}, we arrive at the following 
equation for the pressure field 
\begin{eqnarray}
&&p_{n-1,m}+p_{n+1,m} +p_{n,m-1} +p_{n,m+1}
\nonumber \\      
&=&4\cos{\left(\omega\frac{d }{c_0}\right)}p_{n,m}. 
\label{eigenvalue square}
\end{eqnarray}
To derive the dispersion relation, we seek solutions in the form of Bloch waves
\begin{equation}
p_{n, m}=p_0 e^{\mathrm{i} \mathbf{q} \cdot\mathbf{R}_{n, m}}=p_0 e^{\mathrm{i} q_{x} n d} e^{\mathrm{i} q_{y} m d}, \label{bloch square}
\end{equation}
where $\mathbf{R}_{n,m}=(n,m)d$, $\mathbf{q}=(q_x,q_y)$ is the lattice vector and the Bloch wavenumber respectively, while $q_x$ and $q_y$ are the components of the Bloch wavenumber along the directions $x$ and $y$ of the first Brillouin zone. By substituting the periodic wave solution in Eq.\,\eqref{eigenvalue square} we obtain the dispersion relation of the square lattice
\begin{align}
\cos \left(q_x d\right)+\cos \left(q_y d\right)
=2 \cos{\left(\omega\frac{d}{c_0}\right)}.
\label{dispersion square}
\end{align}
\begin{figure}[tbp!]
\begin{center}
\includegraphics[width=0.5\textwidth]{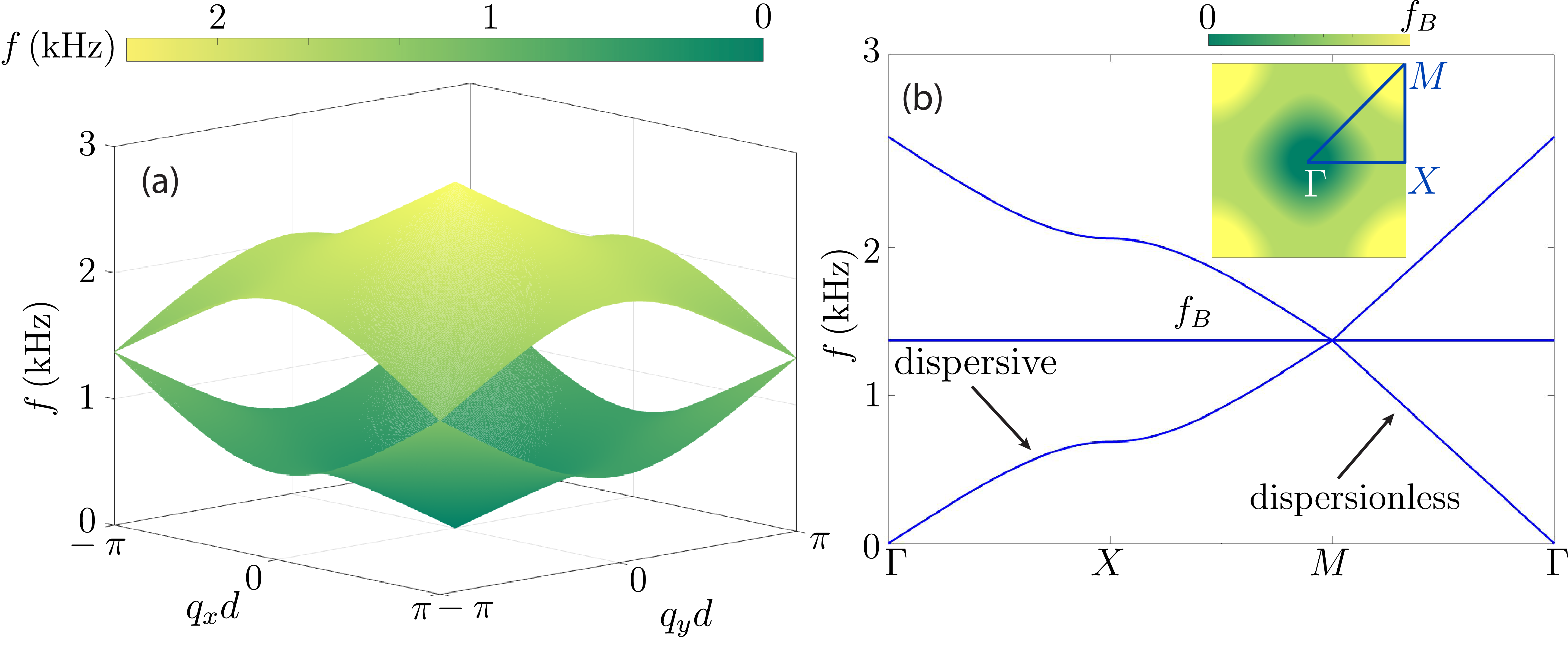} 
 \caption{Band structure of the square network given by \eqref{dispersion square}. (a) Surface of the dispersion relation of the acoustic network. (b) The dispersion relation of the square network along the high-symmetry directions of the first Brillouin zone. The inset shows the first Brillouin zone, while the triangle $\Gamma XM$ denotes its irreducible  representation. The flat band at $f=f_B$, is an exact solution of the 1D Helmholtz equation for the square network--see Eq.\eqref{junctions}-- and corresponds to the resonance $\lambda/2$ of the unit cell, which is also the Bragg frequency.  \label{dispersion brillouin}}
\end{center}
\end{figure}
 The first Brillouin zone of the square lattice is defined as $q_x,q_y\in\,[-\pi/d,\pi/d]$ as illustrated in Fig.~\ref{dispersion brillouin}. Furthermore, the reciprocal lattice exhibits the 8 point-group symmetries of the square, which leaves invariant three distinct points inside the first Brillouin zone. 
 These are known as high symmetry points, $\Gamma=(0,0)$, $X=(\pi/d,0)$ and $M=(\pi/d,\pi/d)$ \cite{Brillouin1946,ashcroft1976}. Consequently, the irreducible representation of the first Brillouin zone are the directions $\Gamma X$, $XM$ and $M\Gamma$, which define the triangle $\Gamma X M$ as presented in Fig.\ref{dispersion brillouin}(b). Notice that the triangle $\Gamma X M$ is the 1/8 of the first Brillouin zone, due to the number of its 8 point-group symmetries.

The above analysis and as it becomes apparent  by looking
at  Fig.\ref{dispersion brillouin}(b),
the dispersion of the square network is highly anisotropic, featuring significant differences between 
the three 
directions. Most notably, the propagation along the horizontal $\Gamma X$ direction of the network features \textit{strong} dispersion, while propagation along the direction $\Gamma M$ is \textit{dispersionless}. The latter is a consequence of in phase multiple scattering, due to the square symmetry of the network; once the lattice distance along the $x$ and $y$ directions become different, the multiple scattering is no longer in phase, and propagation along the diagonal $\Gamma M$ becomes dispersive. 

Finally, we note the existence a flat band around $f=f_B$\, which corresponds to the resonance $\lambda/2$ of the unit cell. This flat band is an exact solution of the 1D Helmholtz equation for the square network --see Eq.\,\eqref{junctions}--and 
corresponds to the Bragg frequency $f_B=c_0/2d$.


\begin{figure}[tbp]
\begin{center}
\includegraphics[width=0.5\textwidth]{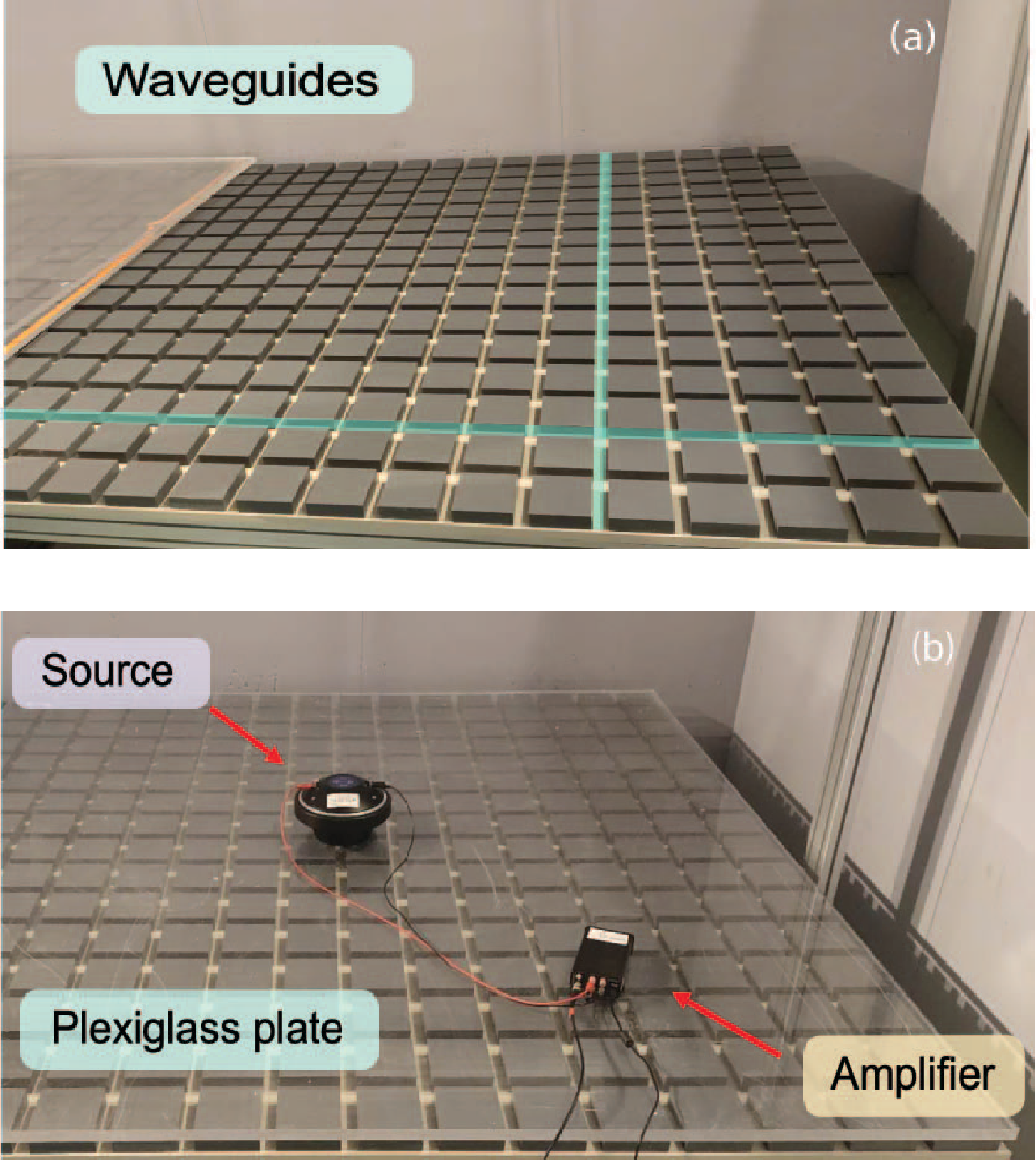}
 \caption{Experimental setup of the square network. \textbf{(a)} A square acoustic network of 15$\times$15 unit cells. A periodic arrangement of square obstacles in square lattice substrate creates waveguide channels, which are depicted in cyan color. \textbf{(b)} Plexiglass plate placed on top of the square obstacles, an acoustic source (compression chamber) placed in the center of the network and an amplifier.} \label{experimental setup square}
\end{center}
\end{figure}

\subsection{Long-wavelength regime}
The dispersive properties 
of the square network, 
can be better  understood by representing the dispersion relation \eqref{dispersion square}  
in an approximate 
polynomial form.
To do this, first we express the Bloch wavenumber in polar coordinates, with ${q=(q_x^2 +q_y^2)^{1/2}}$ and $\theta=\arctan(q_y/q_x)$. 
Then, upon Taylor expanding the dispersion relation 
for long wavelengths ($qd \ll 1$) and low frequencies $(\omega d/c_0\ll 1$), we obtain the following expression
\begin{eqnarray}
\omega(qd) \approx\frac{c_0}{\sqrt{2}d}\left\{q d-\frac{1}{96}\left[1+\cos (4 \theta)\right](q d)^3\right\}. \label{Taylor}
\end{eqnarray}
As seen from Eq.\,\eqref{Taylor}, along the direction $\Gamma M$ (corresponding to $\theta=\pi/4$),  
$\omega(qd) \approx (c_0/\sqrt{2}d)qd$, 
i.e., there is no dispersion. This does not occur along the $\Gamma X$ direction (corresponding to $\theta=0$), where the dispersion coefficient of the $(qd)^3$ term becomes maximum.
Hence, the dispersion relation exhibits different dispersive behaviors, depending on the direction of propagation. To verify the interesting anisotropic behavior we next corroborate our analytical prediction with numerical and experimental results.
\subsection{Experimental validation}
The experimental realization of the 2D square acoustic network is achieved through the periodic arrangement of rigid squares on a square lattice configuration
with lattice constant $d=12.5$\,cm as depicted in Fig.\,\ref{experimental setup square}(a). The rigid square arrangement creates a uniform waveguide network with cross-section ${S_w=2\times2.5\,\text{cm}^2}$ and lattice distance $d$. A plexiglass plate is positioned on top of the squares 
as seen Fig.\,(\ref{experimental setup square})(b). A hole is drilled in the plexiglass plate at the center of the network, where an acoustic source (a compression chamber) is positioned above the opening. A waveform generator and an amplifier are then used to excite sinusoidal and impulsive acoustic waves propagating through the network. Finally, we use a portable microphone (GRAS 40BP) that we can insert inside the network to measure the pressure field at different positions. The total length of the system is $N\times N=15\times15$ unit cells.

\begin{figure}[tbp!]
\begin{center}
\includegraphics[width=0.47\textwidth]{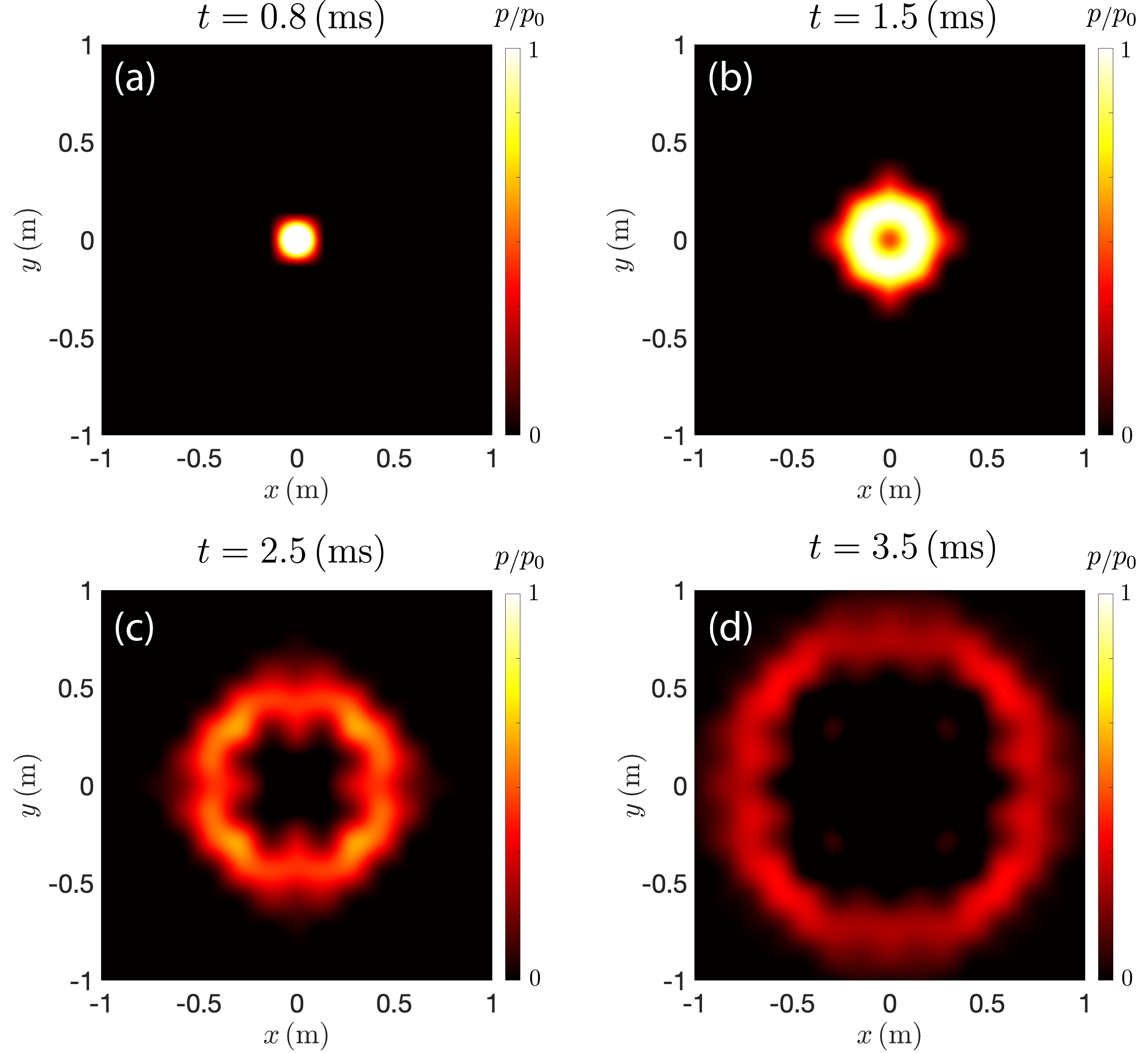} 
 \caption{Contour plots depicting 2D pressure snapshots of cylindrical pulses in the network. Snapshots of the pressure field at times ${t=}{[0.8, 1.5, 2.5, 3.5]}$\,(ms). } \label{snapshots}
\end{center}
\end{figure}

Our first goal is to experimentally evaluate the dispersion relation of the network. 
As an excitation signal, we use a square pulse with spectral width $f_0=500$\,(Hz) and measure the pressure at each junction of the network with our portable microphone. From these measurements we obtain  snapshots of the pressure field, which are normalized by $p_0=500$\,(Pa)--the pressure measured below the compression chamber--, at times $t=[0.8, 1.5, 2.5, 3.5]$\,(ms) presented in Fig.\,\ref{snapshots}. From panels (a-d) one can observe the formation of a cylindrical-like wave propagating through the network. It is readily seen, in  particular, that 
the wave amplitude 
depends on the angle, 
with each maximum lying along the direction $\theta=\pi/4$, for which the dispersion vanishes.

To obtain 
the dispersion relation, we record the temporal signal at each node of the network (i.e., at every junction point) over a time interval long enough for the wave to undergo at least two reflections at the network boundaries. The boundaries are open (Dirichlet boundary conditions, $p = 0$), so the wave reflects, travels back toward the interior, and reaches the boundaries again. This procedure ensures that the recorded signal contains a sufficiently broad range of frequencies, allowing the relevant portion of the dispersion relation to be accurately resolved. The recorded temporal signals are then arranged in a $15\times15$ matrix, where each element has a time dimension.

\begin{figure}[tbp]
\begin{center}
\includegraphics[width=0.46\textwidth]{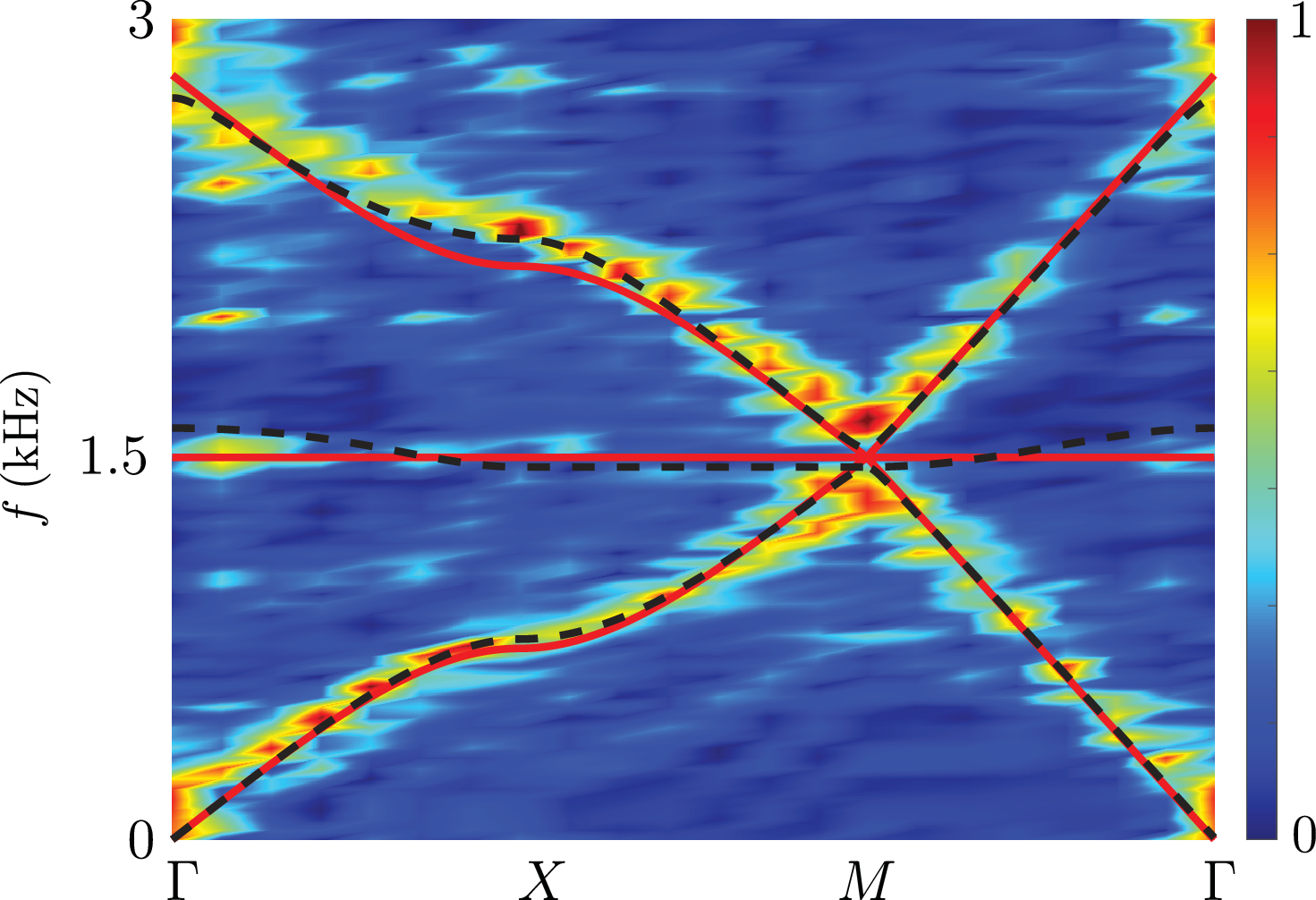} 
\caption{Experimental validation of the dispersion relation. The dispersion relation along the directions of high symmetry of the first irreducible Brillouin zone obtained through the TMM expression \eqref{dispersion square} (red curve), 3D finite element simulations (FEM) (black dashed curve) and experiments (colormap). \label{dispersion experiment}}
\end{center}
\end{figure}

Next, a fast Fourier transform (with respect to time) is performed on each of the matrix's temporal signals, while a 2D fast Fourier transform is performed in space. The resulting matrix represents the dispersion relation of the experimental setup. Consequently, we can 
obtain the dispersion relation along the high-symmetry directions within the first Brillouin zone; for the square lattice the high-symmetry points are $\Gamma$, $X$ and $M$, as defined in the previous section, while the irreducible Brillouin zone is shown in the inset of Fig.\,\ref{dispersion brillouin}(b). Finally, a contour plot of the resulting dispersion relation is shown in Fig.~\ref{dispersion experiment} in the range of $[0,3]$\,kHz, where the first two \textit{propagating} bands are located. 

To verify our experimental results, we compute the dispersion relation, by performing 3D FEM simulations, which are shown in the (black) dotted line and the analytical dispersion relation plotted in the (red) solid line on Fig\,\ref{dispersion experiment}. The analytical method is based on the monomodal approximation; where we consider an effective lattice constant $d^\prime=d+\delta d$ to take into account 2D effects at the intersections. For the experimental structure, we used $\delta d=h_w/2=1.25$\,cm, where $h_w$ is the width of the waveguide.

The three methods are in excellent agreement 
in the entire frequency spectrum  
below the first cutoff frequency of the waveguide segments. Discrepancies occur for the quasi flat band around $f=1.5$\,(kHz) which corresponds to the $\lambda/2$ resonance of the unit cell (which is also the Bragg frequency $f_{\lambda/2}=f_B$). According to the monomodal approximation, the pressure exactly at the center of the node of each unit cell, is zero for all the modes associated with this flat band, while on the same time, these modes exhibit zero group velocity. Due to 3D effects inside each waveguide, the pressure is not exactly zero, hence the numerical flat band is weakly dispersive (i.e exhibits a nonzero group velocity).

Furthermore, the experimental results clearly illustrate that the dispersion of the square network is highly anisotropic and is in good agreement with the analytical predictions of the TMM,  Eq.\,\eqref{dispersion square}. In fact, as long as the width of each waveguide is sufficiently smaller than the lattice distance, i.e., $h_w\ll d$, and frequency below the cutoff of the higher modes of the waveguide, i.e., $kh_w<\pi$, the monomodal approximation is valid.  Finally, for our experimental system, we conclude that if we limit our analysis in the first branch of the dispersion relation, the monomodal approximation is valid and the TMM captures the dispersive behavior of the system with sufficient precision.
\section{Theoretical approach}
As demonstrated in Section\,II, the long-wavelength dispersive behavior of the square lattice is highly anisotropic: it is dispersive along the direction $\Gamma X$, and dispersionless along the direction of $\Gamma M$ (see Figs.\,\ref{dispersion brillouin}, 
\ref{dispersion experiment}). Note that a similar behavior has been reported in a few cases, in 
2D periodic media, lattice structures, and surface waves with periodic bathymetry, fluids with shear flow, where the anisotropy of the dispersion relation leads to pronounced modifications in the form and evolution of propagating wavefields \cite{burz_2d_linear,burz_cylindrical,stepanyants1981,maurel2017revisiting,pham2025homogenized,khusnutdinova_ring,tseluiko2023internal}.  Despite several studies regarding the derivation of an effective PDE that models the anisotropic dispersive properties of 1D and 2D lattices 
\cite{wautier,carta2012dispersive,cornaggia,auffray2015complete,rosi2019continuum}, the aforementioned works do not take into account 
nonlinear phenomena. Here, following the framework established in \cite{vassos_soliton,kinezoula_dark,kinezoula_gap,ioannou2025ring}, we 
will employ the electroacoustic analogue (EA) to model the long-wavelength nonlinear and dispersive characteristics of the network through an effective improved 2D Boussinesq equation.

\subsection{Electroacoustic Analogue--Supercell Transmission line}
The monomodal approximation adopted in Section\,II to characterize the dispersive properties of the square network, allows us to employ a simplified description of our setting, following the analytical framework of \cite{ioannou2025ring}. 
In particular, in each  waveguide segment, either along the $x$- or $y$-direction, the mass conservation (continuity equation) and momentum conservation (Euler equation) take the form, 
\begin{eqnarray}
&&\frac{\partial \varrho}{\partial t}+\frac{\partial}{\partial \nu}(\varrho v_\nu)=0, 
\label{masscons} \\ 
&&\varrho \left(\frac{\partial v_\nu}{\partial t}+v_{\nu}\frac{\partial v_\nu}{\partial \nu}\right) =-\frac{\partial p}{\partial \nu},
\label{momcons}
\end{eqnarray}
where, $\nu=x,y$. Here, $\varrho=\varrho(x,y,t)$ is the density, $v=v(x,y)$ the acoustic velocity, and $p=p(x,y,t)$ is the pressure (all referring to the 
entire network), which are connected via the equation of state $p=p(\varrho,s)$; here $s$ is the entropy, which hereafter is assumed to be constant. 
On the other hand, at each junction, 
to ensure that the coupling between the four connected waveguide segments is properly captured, conservation of mass must be satisfied:
\begin{align}
& 2\frac{\partial \varrho}{\partial t}+\frac{\partial}{\partial x}(\varrho v_x)+\frac{\partial}{\partial y}(\varrho v_y) =0 \label{mass fluid}.
\end{align}
Considering solutions on top of the equilibrium state defined by the density of air $\varrho_0$ and atmospheric pressure $p_{\text{atm}}$, we will make use of the substitutions 
$\varrho\to\varrho_0+\varrho$ and $p\to p_{\text{atm}}+p$. Furthermore, as long as the monomodal approximation is valid, we assume that there is only one velocity component for each waveguide segment, namely $v_x=v_x(x,t)$ and $v_y=v_y(y,t)$, 
the velocity components for a waveguide segment along the $x$- and $y$-directions respectively 
--see Fig.\,\ref{sketch square}(a). 

Next, we employ the quadratic approximation of the equation of state \cite{hamilton,enflo,rudenko}, according to which the density is expressed as the leading-order terms of the Taylor expansion of the pressure, namely
\begin{equation}
\varrho\approx\frac{p}{c_0^2}-\frac{\gamma-1}{2\varrho_0c_0^4} p^2, \label{quadratic}
\end{equation}
where $\gamma$ is the specific heat ratio. We then substitute 
Eq.~(\ref{quadratic}) 
into Eq.~\eqref{masscons}-\eqref{momcons} and keeping only quadratic nonlinear terms in pressure  \cite{hamilton,enflo}, we obtain
\begin{align}
&\frac{1}{c_0^2}\frac{\partial p}{\partial t}
-\frac{\beta_0}{\varrho_0c_0^4}\frac{\partial \left(p^2\right)}{\partial t}
+\varrho_0 \frac{\partial  v_\nu}{\partial \nu}=0, \label{mass simplified waveguide square}  \\
&\varrho_0\frac{\partial v_{\nu}}{\partial  t} =-\frac{\partial p}{\partial \nu}, \label{momentum simplified square}   
\end{align}
which constitute the simplified mass conservation (continuity) and momentum conservation (Euler) equations, respectively, in each of the waveguide segments.

In addition, using Eq.~(\ref{quadratic}), we also approximate  
the density in Eqs.~(\ref{mass fluid}),  
keeping only quadratic nonlinear terms, and obtain the simplified mass conservation for the junction 
\begin{eqnarray}
\frac{2}{c_0^2}\frac{\partial p}{\partial t}
-\frac{2\beta_0}{\varrho_0c_0^4}\frac{\partial \left(p^2\right)}{\partial t}+
\varrho_0 \left(\frac{\partial  v_x}{\partial x}
+\frac{\partial v_y}{\partial y}\right) = 0.
\label{mass simplified square} 
\end{eqnarray}
Direct analytical treatment of the simplified conservation laws (\ref{mass simplified waveguide square}–\ref{mass simplified square}) remains challenging. We therefore adopt the electroacoustic (EA) approach, whereby the conservation laws (\ref{mass simplified waveguide square}–\ref{mass simplified square}) are discretized and recast as an equivalent 2D electrical transmission-line network. Following the methodology developed for square lattices with Helmholtz resonators \cite{ioannou2025ring}, we introduce a refined discretization (supercell) that enables the anisotropic dispersion of the lattice to be accurately captured over a broad frequency range. Keeping the leading-order dispersive and nonlinear effects, the resulting transmission-line model takes the form of the following differential-difference equation (DDE) for the pressure field $(p_{n,m})$
\begin{align}
   &\frac{d^2}{dt^2}p_{n,m}-\frac{c_0^2}{2d^2}\delta_{n,m}^2p_{n,m}+\alpha_{\tilde{N}}\frac{d^2}{c_0^2}\frac{d^4}{dt^4}p_{n,m}=b\frac{d^2}{dt^2}\left(p_{n,m}^2\right),\label{DDE 4th square}
\end{align}
where $\alpha_{\tilde{N}}$ is the dispersion coefficient associated with the 4th-order time derivative, with a distinct value at each supercell order $\tilde{N}$, where $\tilde{N}$ is the number of discrete points for each waveguide segment (for more details see Appendix \ref{appendix electroacoustic}).

 Note that as the supercell order $\tilde{N}$ increases, the dispersion coefficient $\alpha_{\tilde{N}}$ is better approximated; in Appendix\,\ref{appendix electroacoustic} the values of the coefficient are presented up to $\tilde{N}=10$.
Next, the dispersion relation of the EA can be derived upon considering small amplitude plane wave solutions of Eq.~(\ref{DDE 4th square})
\begin{equation}
p_{n,m}(t)=p_0\mathrm{e}^{\mathrm{i}(d\mathbf{q}\cdot\mathbf{n}-\omega t)}+{\rm c.c.},  \label{plane wave}
\end{equation}
where $p_0\ll p_{atm}$, $d\mathbf{q}=(dq_n,dq_m)$, and $\mathbf{n}=(n,m)$. This leads to the dispersion relation
\begin{equation}
    \alpha_{\tilde{N}}\frac{d^2}{c_0^2}\omega^4-\omega^2+\frac{2c_0^2}{d^2}\left[\sin^2{\left(\frac{q_nd}{2}\right)}+\sin^2{\left(\frac{q_md}{2}\right)}\right]=0,\label{discrete dispersion square}
\end{equation}
which, in the long-wavelength approximation, i.e., for ${q_nd}{\ll1}$, $q_md\ll1$, 
takes the form
\begin{align}
&\alpha_{\tilde{N}}\frac{d^2}{c_0^2}\omega^4-\omega^2+\frac{c_0^2}{d^2}\Bigg\{\frac{1}{2}\Big[\left(q_nd\right)^2+\left(q_md\right)^2\nonumber\\
     &-\frac{1}{12}\left(q_nd\right)^4-\frac{1}{12}\left(q_md\right)^4\Big]\Bigg\}=0\label{discrete longwavelength}.
\end{align}
To compare with the TMM analytical results we  Taylor expand \eqref{dispersion square} for long wavelengths, 
$q_x d \ll 1$ and $q_y d \ll 1$, and low frequencies, 
$\omega d/c_0\ll1$, to obtain the following expression
\begin{align}
   &\frac{1}{12}\frac{d^4}{c_0^4}\omega^4- \frac{d^2}{c_0^2}\omega^2+\frac{1}{2}\Big[\left(q_xd\right)^2+\left(q_yd\right)^2\nonumber\\
   &-\frac{1}{12}\left(q_xd\right)^4-\frac{1}{12}\left(q_yd\right)^4\Big]=0 \label{square longwavelength}.
\end{align}
Comparing Eq.\,\eqref{discrete longwavelength}
and Eq.\,\eqref{square longwavelength} we can deduce that as $\tilde{N}\to\infty$, we obtain $\alpha_{\tilde{N}}=1/12$. The relative error between the asymptotic value and the value obtained for $\tilde{N}=10$, is $\delta_{10}=0.01$ (see Appendix\,\ref{appendix electroacoustic}), which is sufficient to model the dispersive 
behavior of the network. In fact, in Appendix\,\ref{appendix electroacoustic}, the relative error is found to be $\delta_N\approx1/\tilde{N}$, and thus the method has an accuracy of $\mathcal{O}(d^2/\tilde{N}^2)$. Hence, as long as the supercell order $\tilde{N}$ is sufficiently high, we can use the \textit{improved} EA to accurately model the long-wavelength and low-frequency behavior of the square network, while also taking into account nonlinear effects in the model, which is one of the main benefits of this scheme.

\subsection{Continuum approximation--2D Improved Boussinesq}

Having established the improved EA approach and the simplified DDE~\eqref{DDE 4th square} describing the long-wavelength and low-frequency dynamics of the network, we will now employ the continuum approximation. 
This 
leads to an effective PDE for the pressure field in the network, namely  
an improved 
2D Boussinesq equation which, in turn, can be reduced  
to the cKdV equation, with an interesting 
feature: the dispersion coefficient 
depends on the propagation angle $\theta$. Notably, for $\theta = \pi/4$, the dispersion coefficient vanishes, and the governing equation reduces to the generalized Burgers equation in cylindrical geometry, which supports cylindrical shock wave solutions \cite{enflo,enflo_N,hamilton}. Thus, the resulting model predicts that the acoustic waveguide network can support cylindrical wave solutions, whose waveform along different directions of propagation, and amplitude-width relation explicitly depend on the propagation angle. 

To be more specific, in the long-wavelength and low-frequency regime, and below the Bragg frequency, ${\omega}{<\omega_B}$,  
the pressure $p_{n,m}(t)$ can be approximated by a continuum variable, i.e., ${p_{n,m}(t)\approx}{ p(x,y,t)}$, where $x=nd$, $y=md$. We approximate the Laplacian by taking into account 
4th-order derivatives
\begin{align}
\!\!\!\!\delta^2_{n,m}p_{n,m}\approx d^{2} \left(\frac{\partial^{2} p}{\partial x^{2}}+\frac{\partial^{2} p}{\partial y^{2}}\right)+\frac{d^{4}}{12}\left( \frac{\partial^{4} p}{\partial x^{4}}+\frac{\partial^{4} p}{\partial y^{4}}\right),
\end{align}
which account for the anisotropy of the dispersion relation \cite{wautier}. This way, the DDE~Eq. \eqref{DDE 4th square} is reduced to the following 
2D Boussinesq equation
\begin{align}
&p_{tt}-c^2\Delta p+\left(\frac{d}{c_0}\right)^2\left[\alpha_{\tilde{N}}p_{tttt}-\frac{c_0^4}{24}\left(p_{xxxx}+p_{yyyy}\right)\right]\nonumber\\
&-b(p^2)_{tt}=0,
\label{Boussinesq square}
\end{align}
where $\Delta \equiv \partial_x^2+\partial_y^2$ is the Laplacian. Although the form of Eq.~\eqref{Boussinesq square} is similar to that obtained in \cite{ioannou2025ring}, here 
the 4th-order  spatial derivative terms  
are also included. 
These terms arise due to the square symmetry of the network \cite{wautier,stepanyants1981}, and were not taken into account in \cite{ioannou2025ring} due to the strong long-wavelength effect of the HRs.
%
The linear dispersion relation of (\ref{Boussinesq improved square}), which can be found by considering  
small-amplitude plane wave solutions ${\propto \mathrm{e}^{\mathrm{i}(\mathbf{k}\cdot\mathbf{r}-\omega t)}+{\rm c.c.}}$ (with $p_0\ll 1$)
%
reads
\begin{align}
\omega(kd,\theta)=&\pm\frac{c_0}{d}\frac{1}{\sqrt{2\alpha_{\tilde{N}}}}\Bigg\{1-\Bigg[1-4\alpha_{\tilde{N}}\left(\frac{c}{c_0}\right)^2(kd)^2\nonumber\\
&+\alpha_{\tilde{N}}\alpha_x^\prime\left[3+\cos{(4\theta)}\right]\left(\frac{c}{c_0}\right)^4(kd)^4\Bigg]^{1/2}\Bigg\}^{1/2},
\label{dispersion supercell}
\end{align}
where $k=(k_x^2+k_y^2)^{1/2}$, and $\pm$ signs correspond to outgoing- and ingoing waves. It is worth noting that in Eq.\, \eqref{dispersion supercell} we kept only the solutions corresponding to the first branch of the dispersion relation.

\begin{figure}[tb!]
\begin{center}
\includegraphics[width=0.45\textwidth]{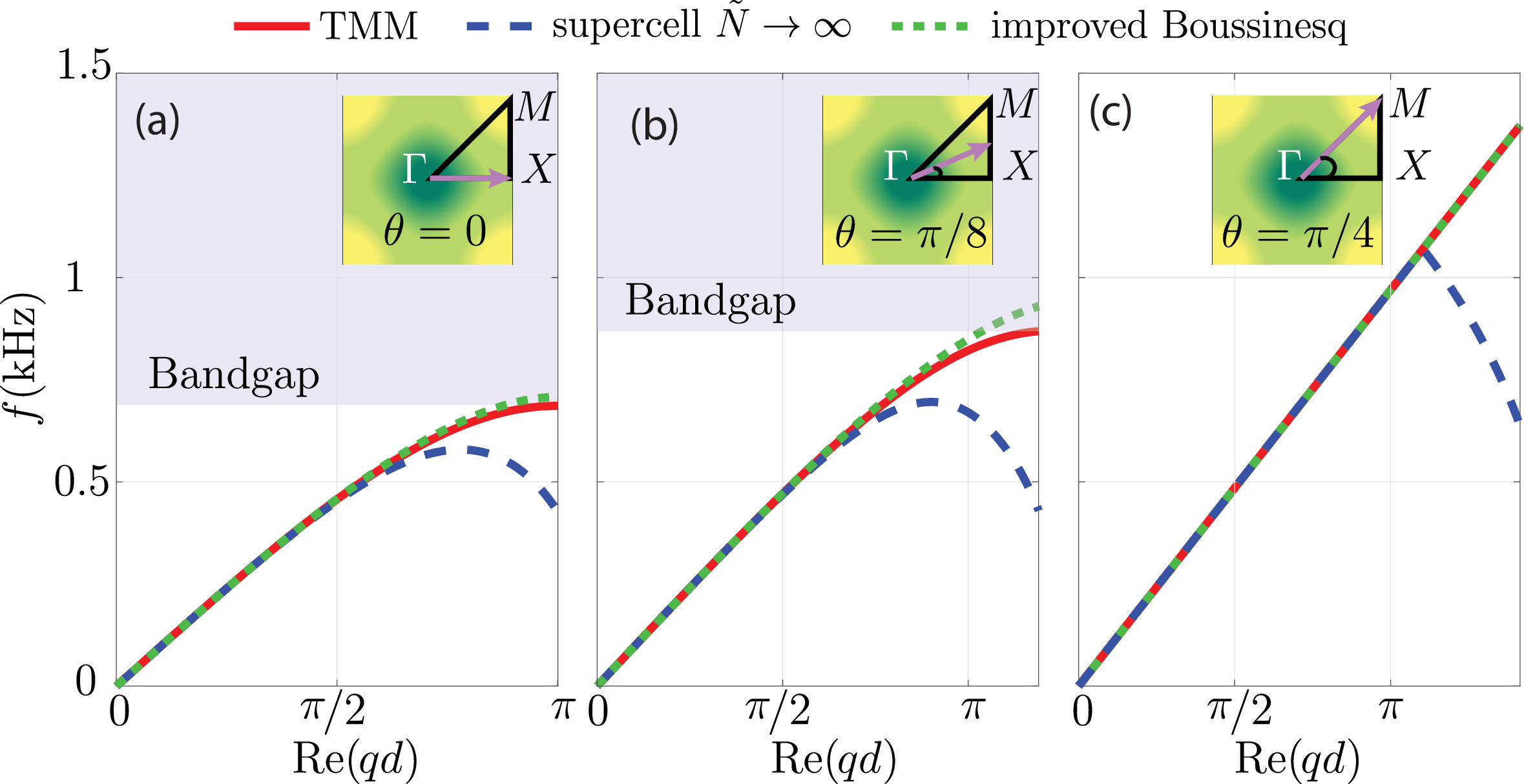} 
 \caption{Dispersion relation of square network, along different direction of propagation $\theta$. The solid (red) line depicts the TMM dispersion relation given by \eqref{dispersion square}, the dashed (blue) line depicts the dispersion relation of the continuum approximation of the supercell transmission line taking $\tilde{N}\to\infty$ \eqref{dispersion supercell}, and the dotted (cyan) line corresponds to the optimized Boussinesq (with $\alpha_t=0$) dispersion relation of the improved Boussinesq Eq.\,\eqref{dispersion improved square}. Panels (a-c) depict the dispersion for three different directions $\theta=[0,\pi/8,\pi/4]$ respectively. The pale purple colour depicts the bandgap (present in the first two cases), while inside the inset, a (purple) vector illustrates the direction in the first Brillouin zone.
  \label{boussinesq dispersion}}
\end{center}
\end{figure}

A comparison between the dispersion relation obtained from the TMM approach, Eq.\,\eqref{dispersion square} 
[solid (red) curve], and that predicted by the Boussinesq equation,  
Eq.~\eqref{dispersion supercell}  
[dashed (blue) curve], is presented in Fig.\,\ref{boussinesq dispersion} for 
angles 
$\theta = [0, \pi/8, \pi/4]$, in panels (a–c), respectively. The purple regions indicate the bandgaps along each direction
%
(no bandgap 
occurs for $\theta = \pi/4$) and, in each inset, a purple arrow marks the corresponding propagation direction in the first Brillouin zone. These three angles are chosen because the square lattice exhibits maximum dispersion for $\theta = 0$, retains dispersion with distinct characteristics at the intermediate angle $\theta = \pi/8$, and becomes dispersionless for $\theta = \pi/4$. As shown in Fig.\,\ref{boussinesq dispersion}, the 
Boussinesq model captures the TMM dispersion curves with good accuracy across all three propagation directions.

For $\theta=0$ and $\theta = \pi/8$, depicted in panels (a) and (b), the agreement is excellent in the long-wavelength regime and gradually deteriorates as the wavelength decreases. As expected, near the edge of the Brillouin zone, the effective model for $\theta = 0$ fails to reproduce the correct dispersive behavior of the acoustic network. For ${\theta =}{\pi/4}$, shown in panel (c), the effective model successfully reproduces the dispersionless behavior of the network; however, at shorter wavelengths it again deviates from the TMM prediction, consistent with typical limitations of Boussinesq-type approximations \cite{Remoissenet,wautier,cornaggia,higher_order,bona,madsen1991new,madsen2003boussinesq,nwogu1993alternative}.

Here, it should be noted that 
the derived Boussinesq  Eq.~\eqref{Boussinesq square} 
is characterized as an \emph{ill-posed} (or ``bad'') Boussinesq equation  \cite{Remoissenet,higher_order,higher_order_2}, as it supports 
nonphysical unstable solutions, featuring exponential growth at short wavelengths (for sufficiently large $kd$, the frequency $\omega$ in \eqref{dispersion supercell} becomes complex). Nevertheless, in weakly dispersive systems, this problem is commonly mitigated by replacing the fourth-order spatial derivatives --and in the present case also the fourth-order temporal derivative-- with the mixed fourth-order derivative \cite{Remoissenet}. 
Alternatively, one may use a suitable (often fitted) linear combination of fourth-order spatial, temporal, and mixed derivatives, derived using the leading-order wave equation \cite{wautier,cornaggia}. 
This problem also occurs in 2D Boussinesq-type equations \cite{wautier}, where the same approach is used 
to regularize the short-wavelength behavior of the model.


Here, we follow the approach of Refs.~\cite{wautier,cornaggia} and regularize 
the ill-posed Boussinesq equation as follows.
First, introducing the transformations 
%
%
\begin{align}
p \to \frac{p}{p_0}, \quad t\to\frac{c_0}{d}t,\quad x\to\frac{x}{d},\quad y\to\frac{y}{d}. \label{normalized square}
\end{align}
where $p_0$ denotes the wave amplitude, we express 
Eq.\,\eqref{Boussinesq square} 
in the following normalized form:
 \begin{align}
&p_{tt}-\left(\frac{c}{c_0}\right)^2\Delta p+\alpha_{\tilde{N}}p_{tttt}-\frac{1}{24}\left(p_{xxxx}+p_{yyyy}\right)\nonumber\\
&-\frac{b}{p_0}(p^2)_{tt}=0.\label{Boussinesq normalised}
\end{align}
It is now clear that the coefficients of the dispersive and nonlinear terms are of  order $\mathcal{O}(\varepsilon)$, since ${\alpha_{\tilde{N}}}{=1/12,\,\,1/24\ll1}$, 
and $b/p_0\sim\varepsilon$. Consequently, to leading order $\mathcal{O}(\varepsilon^0)$, 
we obtain the 2D wave equation, 
as in \cite{wautier}
\begin{equation}
p_{tt}-\left(\frac{c}{c_0}\right)^2\Delta p=0, \label{leading order square}
\end{equation}
where $c=c_0/\sqrt{2}$ is the effective speed of sound. By operating with the second order temporal derivative and the Laplacian on the leading order wave equation of Eq.\,\eqref{leading order square}, we obtain the following relations
\begin{align}
p_{tttt}=\left(\frac{c}{c_0}\right)^2\Delta p_{tt}=\left(\frac{c}{c_0}\right)^4\Delta^2p,\ \label{derivatives}
\end{align}
where $\Delta^2=\partial_{xxxx}+2\partial_{xxyy}+\partial_{yyyy}$ is the biharmonic operator in cartesian coordinates. 
%
Next, employing 
Eq.\,\eqref{derivatives}, 
we replace 
the 4th-order temporal derivative of the Boussinesq Eq.\,\eqref{Boussinesq square}, 
and obtain the equation
\begin{equation}
\!\!\!\alpha_{\tilde{N}}p_{tttt}=\alpha_tp_{tttt}+\alpha_m\left(\frac{c}{c_0}\right)^2\Delta p_{tt}-\left(\frac{c}{c_0}\right)^4\alpha_x\Delta^2 p,
\end{equation}
where $\alpha_t$, $\alpha_m$, and $\alpha_x$  are the dispersive coefficients of the temporal, the mixed, and the isotropic spatial derivatives, respectively. By retaining terms up to order $\mathcal{O}(\varepsilon)$, we finally obtain the \textit{improved} Boussinesq equation
\begin{align}
&p_{tt}-\left(\frac{c}{c_0}\right)^2\Delta p+\alpha_tp_{tttt}+\alpha_m\left(\frac{c}{c_0}\right)^2\Delta p_{tt}-\alpha_x\left(\frac{c}{c_0}\right)^4\nonumber\\
&\times\Delta^2 p-\alpha_x^\prime \left(\frac{c}{c_0}\right)^4\left(p_{xxxx}+p_{yyyy}\right)=\tilde{b}(p^2)_{tt},\label{Boussinesq improved square}
\end{align}
where $\alpha_x^\prime$ is the coefficient of the anisotropic 4th-order spatial derivative. 

The inclusion of  
more dispersion coefficients, 
not only ensures that the spurious short-wavelength instability 
of the ill-posed Boussinesq 
\eqref{Boussinesq normalised} ceases to exist, but also enables a better agreement with  
the 
dispersion relation of the TMM, Eq.\,\eqref{dispersion square}. An \textit{optimal} set of the aforementioned coefficients can be determined by applying the proper fitting constraints \cite{wautier,cornaggia} (see 
Appendix\,\ref{appendix optimal dispersion} for details). 

The linear dispersion relation  
of Eq.~(\ref{Boussinesq improved square}) reads 
\begin{align}
&\omega(kd,\theta)=\pm\frac{c_0}{d}\frac{1}{\sqrt{2\alpha_t}}\Bigg\{1-\alpha_m\left(\frac{c}{c_0}\right)^2(kd)^2-\nonumber\\
&\Big[1-2\left(\alpha_m+2\alpha_t\right)\left(\frac{c}{c_0}\right)^2(kd)^2+\big(\alpha_m^2+4\alpha_t\alpha_x\nonumber\\
&+\alpha_t\alpha_x^\prime\left[3+\cos{(4\theta)}\right]\big)\left(\frac{c}{c_0}\right)^4(kd)^4\Big]^{1/2}\Bigg\}^{1/2}, \label{dispersion improved square}
\end{align}
where 
we have used the original variables, 
as in Eq.\,\eqref{boussinesq dispersion} 
in the previous Section.

In Fig.\,\ref{boussinesq dispersion}, we compare the optimal effective dispersion relation Eq.\,\eqref{dispersion improved square} [dotted (cyan) curve] with the TMM prediction, Eq.\,\eqref{dispersion square} [solid (red) curve].
We find an excellent agreement between the two, over a broad range of wavelengths, including the short-wavelength regime. 
Only small deviations near the edge of the Brillouin zone for 
$\theta=0$ and $\theta=\pi/8$ is observed [see 
panels (a) and (b), respectively], while 
for $\theta=\pi/4$ pertinent curves are practically identical 
[panel (c)].
Note that the optimized model is obtained by setting $\alpha_t = 0$ (see also Appendix\,\ref{appendix optimal dispersion}).

According to the above analysis, the resulting effective improved Boussinesq equation --expressed in the original variables, as Eq.\,\eqref{Boussinesq square}--reads 
\begin{align}
&p_{tt}-c^2\Delta p+\left(\frac{d}{c_0}\right)^2\Big[\alpha_m c^2\Delta p_{tt}-\alpha_x c^4\Delta^2 p\nonumber\\
&-\alpha_x^\prime  c^4\left(p_{xxxx}+p_{yyyy}\right)\Big]-b(p^2)_{tt}=0.\label{Boussinesq improved square2}
\end{align}
%
Note that this equation is a 
generalization of the effective 
1D Boussinesq model, previously 
derived for an acoustic periodic waveguide \cite{sougleridis2023acoustic}.

\subsection{Cylindrical KdV Equation}
We now seek small-amplitude cylindrical solutions of Eq.~(\ref{Boussinesq improved square2}); the Laplacian the biharmonic operator and the 4th order spatial derivatives in $x$ and $y$ can be expressed in polar coordinates $(r,\theta)$ as found in Appendix \ref{appendix derivatives}, where $r=\sqrt{x^2+y^2}$ and and ${\theta}{=\arctan{(y/x)}}$ are the radial and  
angular coordinates. 

Having expressed 
the improved Boussinesq Eq.\,\eqref{Boussinesq improved square2} in 
polar coordinates, it becomes apparent that the anisotropy of the square lattice is induced by the 4th-order spatial derivatives. Notice that for radially symmetric initial/boundary conditions, the remaining of the 4th-order dispersive terms support a radially symmetric solution, while the spatial 4th-order derivatives break the radial symmetry, due to the anisotropy. 

We continue by seeking radially symmetric solutions of Eq.~\eqref{Boussinesq improved square2} in the form of the following asymptotic expansion
\begin{equation}
    p=\varepsilon p_1+\varepsilon^{2} p_2+\varepsilon^{3} p_3+\cdots, \label{series2}
\end{equation}
where $0<\varepsilon \ll 1$ is a formal small parameter, and $p_i$ ($i=1,2,\ldots$) are unknown functions depending on the slow variables
\begin{equation}
\!\!\!T=\varepsilon^{1/2}\left(\frac{r-r_0}{c}-t\right),\quad R=\varepsilon^{3/2}(r-r_0), \label{slow scales square cylindrical}
\end{equation}
where $r_0$ is the initial radius of the wave. Substituting Eq.~(\ref{series2}) into Eq.~(\ref{Boussinesq improved square2}), and using the slow variables of Eq.\,\eqref{slow scales square cylindrical}, we obtain identities at orders $\mathcal{O}(\varepsilon)$ and $\mathcal{O}(\varepsilon^2)$, while at order $\mathcal{O}(\varepsilon^3)$, we find the cKdV equation
\begin{equation}
  {p_1}_{R}+\tilde{\alpha}(\theta){p_1}_{TTT}+\tilde{\beta} p_1{p_1}_{T}+\frac{1}{2R} {p_1}=0, \label{ckdv square}
\end{equation}
with 
\begin{align}
     \tilde{\alpha}(\theta)=\frac{1}{2c}\left(\frac{d}{c_0}\right)^2\left[\alpha_x+\frac{3+\cos{(4\theta)}}{4}\alpha_x^{\prime}-\alpha_m \right],\,\,\, \tilde{\beta}=\frac{b}{c},\nonumber
\end{align}
where $\tilde{\alpha}(\theta)$ and $\tilde{\beta}$ are the dispersion and nonlinearity coefficients, respectively.
Equation~\eqref{ckdv square} is the cKdV equation, with 
a dispersion coefficient, $\tilde{\alpha}(\theta)$, depending explicitly on the angle $\theta$; 
this reflects the anisotropy of the square lattice, 
as in \cite{stepanyants1981}.
Consequently, Eq.~\eqref{ckdv square} describes the propagation of high-amplitude, cylindrical-shaped waves whose amplitudes and waveforms depend explicitly on the propagation angle $\theta$, sufficiently far from the source. In particular, as 
will be shown below,
for $\theta \neq \pi/4$, the corresponding solutions 
take the form of solitons with angle-dependent amplitude–width relations, 
while for $\theta \to \pi/4$, where dispersive effects vanish, the equation admits shock wave solutions.

\section{Linear and Nonlinear Cylindrical Waves}
We now study the propagation of linear and nonlinear cylindrical-shaped waves 
in the acoustic network, in the framework of the cKdV (\ref{ckdv square}). We first examine linear waves of low-amplitude, such that the nonlinearity 
in Eq.\,(\ref{ckdv square}) becomes negligible. Then, we study the fully nonlinear version and investigate the dynamics of cylindrical solitons and shock waves, along different directions of propagation.

\subsection{Linear cylindrical waves -- self-similarity and anisotropy}

For sufficiently low pressure
\begin{figure}[tb!]
\begin{center}
\includegraphics[width=0.45\textwidth]{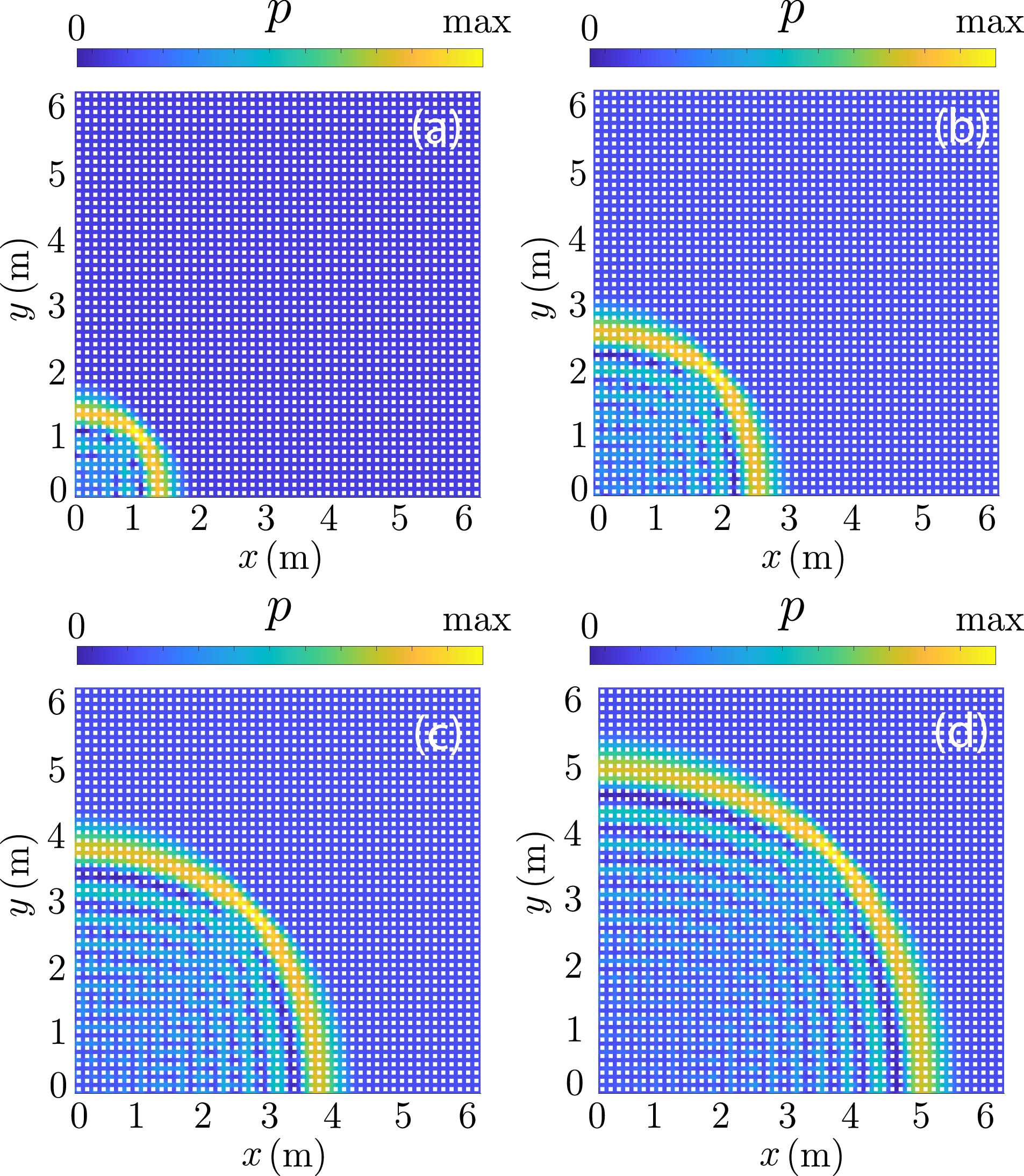}
  \caption{Contour plots of the pressure field inside the waveguide network, in the low-amplitude limit, at times (in ms) $t=10$, $15$, $20$, $25$ in panels (a)-(d) respectively.
  \label{contour linear square}}
  \end{center}
\end{figure}
amplitudes --specifically, for a normalized amplitude $p_0$ on the order of $1~\mathrm{Pa}$--the nonlinear term in Eq.\,\eqref{ckdv square} can be neglected 
and we therefore consider 
its linearized form, namely
\begin{equation}
    p_{1R}+\tilde{\alpha}(\theta)p_{1TTT}+\frac{1}{2R}p_1=0, 
    \label{linearized ckdv square}
\end{equation}
together with the uniform (along $\theta)$ pressure condition imposed at the central node of the network ${p(R=0, T)}{= p_0(T)}$. For $\theta\neq\pi/4$, we may follow  \cite{dorfman_airy,ioannou2025ring} and 
seek for cylindrical self-similar solutions 
 of Eq.\,\eqref{linearized ckdv square}.  
In this case, for each direction of propagation, the solutions preserve their temporal shape while exhibiting spatially dependent amplitudes and/or widths (see Ch.~5 of Ref.~\cite{logan2008}). A self-similar solution of Eq.~\eqref{linearized ckdv square} can therefore be sought through the following ansatz
\begin{equation}
    p_{\text{sim}}(\theta)=\frac{1}{R^w}f(\eta),\quad \eta=\frac{BT}{R^q}, \quad \theta\neq\frac{\pi}{4},\label{self ansatz square}
\end{equation}
where $f(\eta)$ is an unknown function of the similarity coordinate $\eta$ that satisfies homogeneous boundary conditions at infinity, i.e., $f(\eta)\to 0$ as $\eta\to\infty$. The quantities $B$, $w$, and $q$ are parameters to be determined independently, to ensure self-similarity for each direction of propagation. 

Substituting the ansatz \eqref{self ansatz square} into Eq.~\eqref{linearized ckdv square}, we find that 
$w=5/6$ and $B=(3\tilde{\alpha}(\theta))^{-1/3}$, and hence 
the self-similar solution is given by
  \begin{align}
    &p(r,\theta,t)\sim \frac{\hat{p}_0(0)}{(r-r_0)^{-5/6}}\Ai(\eta), \nonumber\\ 
    &\eta =\frac{\frac{r-r_0}{c}-t}{\left[3\tilde{\alpha}(\theta) (r-r_0)\right]^{1/3}},\quad \theta\neq\frac{\pi}{4},
    \label{airy solution square}
\end{align}
where $\hat{p}_0(0)$ is the first term of the Taylor expansion of the Fourier transform of the initial condition $p_0(T)$  and $\Ai(\eta)$ is the Airy function, whose integral form 
reads
\begin{equation}
    \Ai(\eta)=\frac{1}{2\pi}\int_{-\infty}^{+\infty}ds 
    \exp \left[\mathrm{i}\left(s\eta+\frac{s^3}{3}\right)\right]. \label{airy}
\end{equation}
From the asymptotic formulae of the Airy function one can deduce that 
for $T\rightarrow 0$ the solution will vary as $p(T,R,\theta)\sim R^{-5/6}$ for $\theta\neq\pi/4$. The value of $\theta=\pi/4$ corresponds to the direction along the diagonal $\Gamma M$, and the linearized cKdV Eq.\,\eqref{linearized ckdv square} reduces to the radial transport equation, $ p_{1R}+p_1/2R=0,$ for which the solution of the boundary values problem is
\begin{equation}
p_1(R,T)=p(T)\left(\frac{R_0}{R}\right)^{1/2},\quad \theta=\pi/4. \label{transport square}
\end{equation}
Hence, the decay law is 
$p(T,R,\pi/4)\sim R^{-1/2}$.
We note that our result agrees with the calculation of the far field decay of the cylindrical wave equation (see Ch.~7 in \cite{Whitham}).

\subsubsection{
Numerical results -- linear regime}
Next, we compare 
the analytical anisotropic self-similarity predictions 
with numerical simulations. 
%
For the time dependent numerical simulations we directly solve the 2D wave equation in the square network, 
\begin{equation}
p_{tt} - c_0^2 \Delta p = 0, \quad \partial_n p = 0\,\,\, \text{on the walls},\,
\label{2d wave}
\end{equation}
where 
``walls'', denote the boundaries of each waveguide. Equation~\eqref{2d wave} is solved using the transient acoustic module of \textsc{COMSOL Multiphysics}.
This numerical approach 
accurately captures the wave dynamics of the square network in the monomodal approximation, 
and also fully accounts for 2D effects. Consequently, it remains valid over a broader frequency range than the transmission-line approach 
\cite{kinezoula_dark,kinezoula_dark,vassos_soliton,sougleridis2023acoustic,ioannou2025ring}, albeit at a higher computational cost.

 We choose a finite network of $80\times80$ unit cells. The lattice distance is the same as our experimental setup in Section II, $d=12.5$\,cm and the width of the waveguide $h_w=0.5$\,cm, such that 2D effects are negligible.
 
 We consider a boundary condition, at $r=0$, of the Gaussian form  
\begin{equation}
p(t,0)= p_0\exp\left\{-\left[(t-t_0)/2\sigma\right]^{2}\right\}, \label{BC square}
\end{equation} 
where $p_0$ and $\sigma$ are the amplitude and standard deviation. We fix $p_0=1$\,Pa and use 
$\sigma=0.45$\,ms, corresponding to a half-width of $\approx 0.82$~ms.

Contour plots of the resulting pressure field are presented in  Fig.\,\ref{contour linear square} at times $t=[10,\,15,\,20,\,25]$\,ms, in panels~(a-d) respectively. 
It is observed that the width  and amplitude of the pulse vary with the direction of propagation, with its minimum width and maximum amplitude lying in the diagonal $\theta=\pi/4$ 
(dispersionless limit). 
For $\theta \neq\pi/4$, the amplitude of the pulse decays rapidly and becomes wider, with its minimum amplitude and wider duration for $\theta=0$.

\begin{figure}[tb!]
\begin{center}
\includegraphics[width=0.5\textwidth]{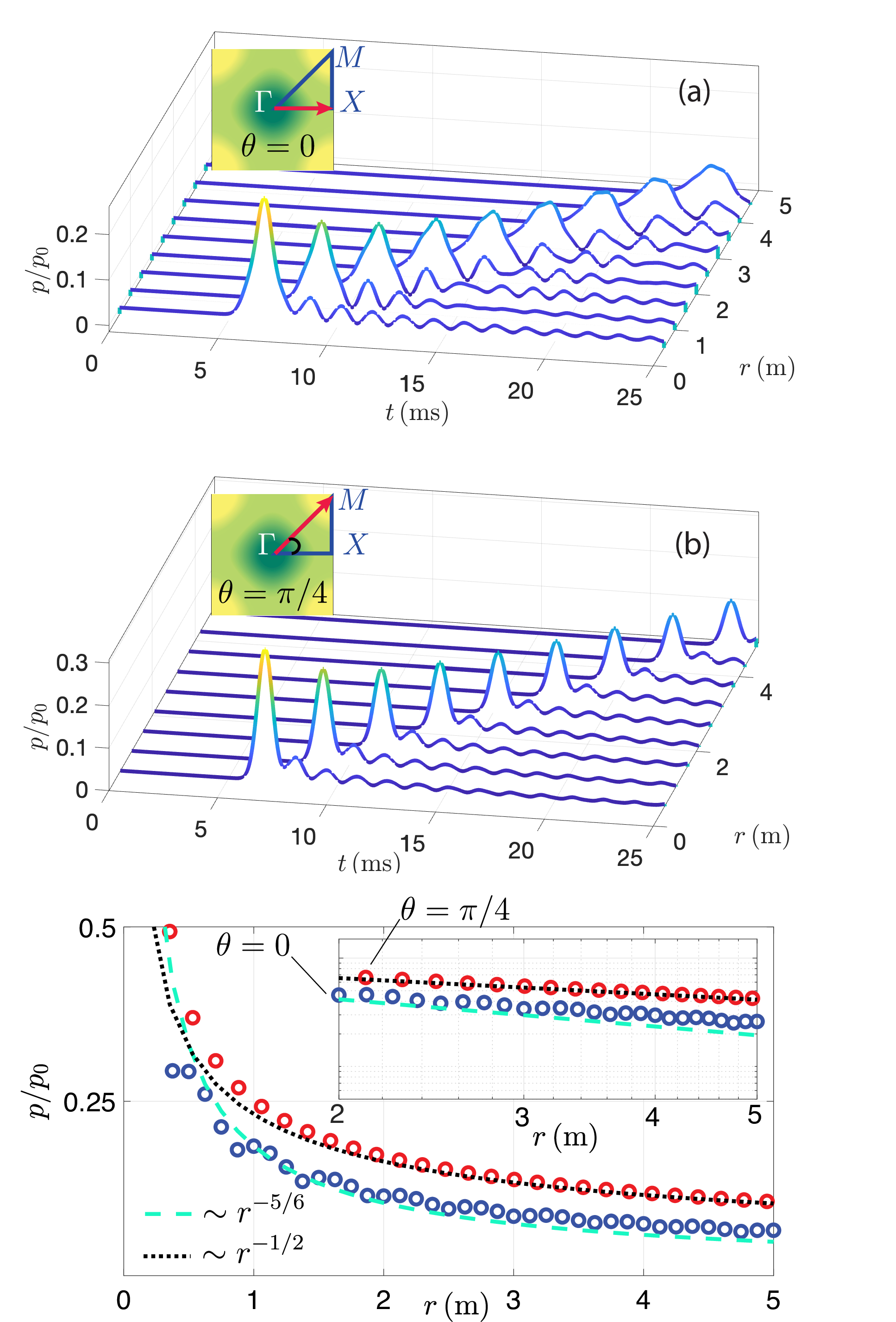} 
 \caption{(a) Evolution of the linear pressure pulse along the direction $\theta=0$ at fixed distances. (b) Same along the direction $\theta=\pi/4$. (c) Amplitude decay along the direction $\Gamma X$ and $\Gamma M$; numerical results are presented in circles for the direction $\theta=0$ (blue) and $\theta=\pi/4$ (red) and the analytical predictions of Eq.\,\eqref{airy solution square} in dashed (cyan) line, and Eq.\,\eqref{transport square} dotted (black). The inset depicts a zoom of the amplitude decay in log-log scales, which is a straight line of slope $-5/6$ for the direction $\theta=0$ and $-1/2$ for $\theta=\pi/4$.
  \label{temporal linear square}}
\end{center}
\end{figure}

The evolution of the low-amplitude pulse along the directions $\theta=0$ and $\theta=\pi/4$ is also depicted in the 3D plot of Fig.~\ref{temporal linear square}, in panels (a) and (b) respectively. For 
$\theta=0$, each of the individual snapshots, shown 
at fixed distances, features an increasing width due to dispersion and rapid amplitude decay, as predicted by Eq.\,\eqref{airy solution square}. In panel (b) the pulse  maintains its width, and significantly slower amplitude decay. as predicted by Eq.\,\eqref{transport square}. 

The amplitude decay along both directions $\theta=0$ and $\theta=\pi/4$ is presented in
Fig.~\ref{temporal linear square}(c). In particular, the numerical results are presented in circles for the direction $\theta=0$ (blue) and $\theta=\pi/4$ (red) and the analytical predictions of Eq.\,\eqref{airy solution square} in dashed (cyan) line, and Eq.\,\eqref{transport square} dotted (black). The inset depicts a zoom of the amplitude decay in log-log scales, which is a straight line of slope $-5/6$ for the direction $\theta=0$ and $-1/2$ for $\theta=\pi/4$. There is a very good agreement, between the theoretical predictions and the numerical results.

The 
different ampliude decays highlight the anisotropic structure of the solution. Finally, we note that, at the end of the simulation, the ratio of the amplitudes 
in the directions $\theta=0$, and $\theta=\pi/4$ is $\approx 0.55$.

\subsection{Nonlinear cylindrical waves -- shock waves and solitons}
We now study the case of high-amplitude waves, where dispersion, nonlinearity, and curvature-induced decay terms of the cKdV 
(\ref{ckdv square}) are of the same order. 

\subsubsection{Cylindrical Solitons}
We first consider  
cylindrical soliton solutions of the cKdV. 
%
Following  
Ref. \cite{KO}, 
an approximate cylindrical soliton 
for $\theta\neq\pi/4$ takes the form
\begin{align}
p_1(T,R,\theta)\approx A(R,\theta)\sech^2{\left[w_0(R_0,\theta)\left(T-\frac{R-R_0}{v(R_0,\theta)} \right)\right]},\label{soliton approximate square}
\end{align}
where $A(R,\theta)=A_0(\theta)\left(R_0/R\right)^{2/3}$ is the spatially-varying soliton amplitude (with $A_0(\theta)$ being the soliton amplitude at the initial radius $R=R_0$ for each direction $\theta$), while the soliton's width $w_0(R,\theta)$ and velocity $v(R,\theta)$ are 
given by
\begin{eqnarray}
w_0(R,\theta)=\left(\frac{A(R,\theta)\beta}{12\tilde{\alpha}(\theta)}\right)^{1/2}, 
\quad 
v(r)=\frac{3}{A(R,\theta)\tilde{\beta}}.
\end{eqnarray} 
Notice that the above soliton  is characterized by 
parameters 
(amplitude, width and velocity) that depend on the angle of propagation, similarly to the cylindrical soliton supported by a square lattice of 
transmission lines 
\cite{stepanyants1981}. 

Equation~(\ref{soliton approximate square}) is expressed in terms of the original variables, $r$, $\theta$ and $t$, as follows, 
\begin{align}
p(r,\theta,t) \approx &\varepsilon A(r,\theta)  
\sech^2 \Big\{ \varepsilon^{1/2}w_0(r,\theta)\nonumber\\
&\times\left[\left(\frac{1}{c}-\varepsilon\frac{1}{v(r,\theta)}\right)\left(r-r_0\right)-t\right]\Big\}. \label{soliton solution square}
\end{align}
In either representation, it becomes clear that the balance of dispersion, nonlinearity, and curvature 
depends on the direction of propagation. The soliton solution is wider for $\theta=0$ and becomes thinner as $\theta$ increases. 
For $\theta\to \pi/4$, the solution ceases to exist
since the dispersion coefficient vanishes. 
In this case, Eq.\,\eqref{ckdv square} reduces to the radial inviscid Burgers equation 
which models the formation of cylindrical shock waves \cite{hamilton,enflo,enflo_N}. 

In what follows, we use the radial inviscid Burgers equation to obtain the solution before the shock formation, as well as the wave breaking distance (i.e., the shock formation distance).


 \begin{figure}[h!]
\begin{center}
\includegraphics[width=0.45\textwidth]{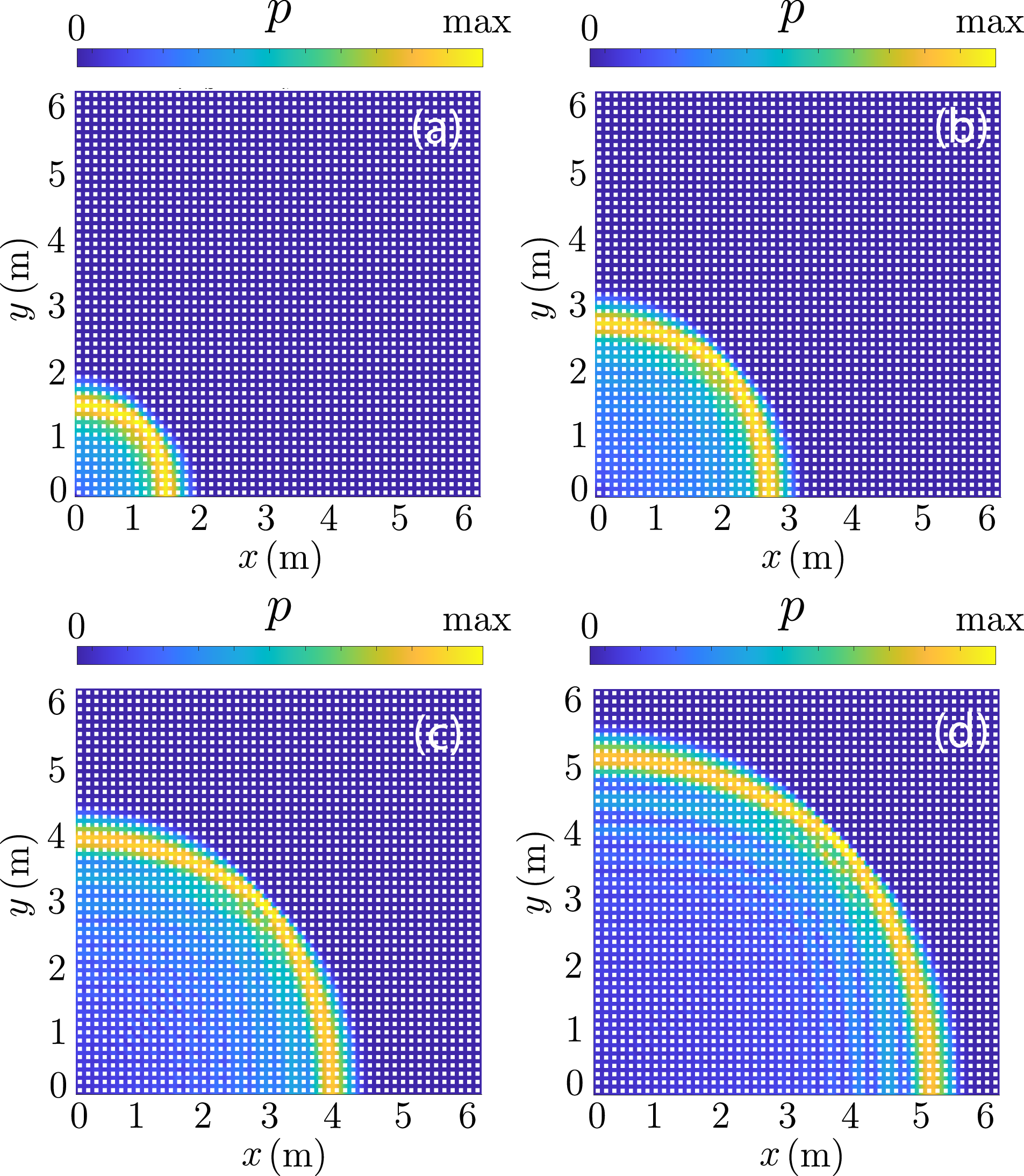}
  \caption{Contour plots of the pressure field inside the waveguide network, in the high-amplitude limit, at times $t=[10,\,15,\,20,\,25]$\,ms in panels (a-d) respectively.
  \label{contour nonlinear square}}
  \end{center}
\end{figure}

\subsubsection{Cylindrical shock waves}
As mentioned above,
in the case of vanishing dispersion 
($\theta=\pi/4$), 
and Eq.\,\eqref{ckdv square} 
reduces to 
\begin{equation}
{p_1}_{R}+\tilde{\beta}p_1{p_1}_{T}+\frac{1}{2R}p_1=0, \label{radial burgers}
\end{equation}
supplemented with the boundary condition ${p_1(R_0,T)=}{g(T)}$. 
Introducing the transformations
\begin{align}
   z=2\left(\sqrt{R}-\sqrt{R_0}\right)\sqrt{R_0},\quad  \phi(T,z)=\left(\frac{R}{R_0}\right)^{1/2}p_1,
\end{align}
we obtain from Eq.\,\eqref{radial burgers}
an inviscid Burgers equation with constant coefficients,
\begin{equation}
\phi_z+\tilde{\beta}\phi\phi_T=0. \label{burgers transform}
\end{equation}
Employing the method of characteristics \cite{Whitham}, we find the implicit solution 
\begin{equation}
    \phi(z,T)=g(T-\tilde{\beta}\phi z), \label{characteristics}
\end{equation}
%
Considering, e.g., a Gaussian boundary condition (see Eq.~\eqref{BC square}) the solution 
\eqref{characteristics} 
remains valid up to a certain distance, referred to as ``breaking distance'', $z_B$; at this distance, a dicontinuity of the solution emerges, i.e., a shock wave is formed (for $z>z_B$ the solution becomes multivalued and, as such, ceases to exist) \cite{Whitham}.  
The breaking distance is given by 
\begin{align}
    z_B=\min_{\xi>0}\left\{-\frac{1}{\tilde{\beta}g^\prime(\xi)}\right\},\quad \text{with} \quad  g^\prime<0,
\end{align}
and for  
$g(T)=\tilde{p_0} \exp{\left[-(T-T_0)^2/(2\tilde{\sigma}^2)\right]}$ (where $\varepsilon \tilde{p}_0=p_0$, and $\tilde{\sigma}=\varepsilon^{1/2}\sigma$) we find that 
\begin{equation}
    z_B=\frac{1.6\tilde{\sigma}}{\tilde{\beta} \tilde{p}_0}.
\end{equation}
This, in turn, leads to 
the 
breaking radius $R_B$, given by
\begin{align}
    R_B=R_0\left(1+\frac{z_B}{2R_0}\right)^2.
\end{align}

\begin{figure}[tbp!]
\begin{center}
\includegraphics[width=0.45\textwidth]{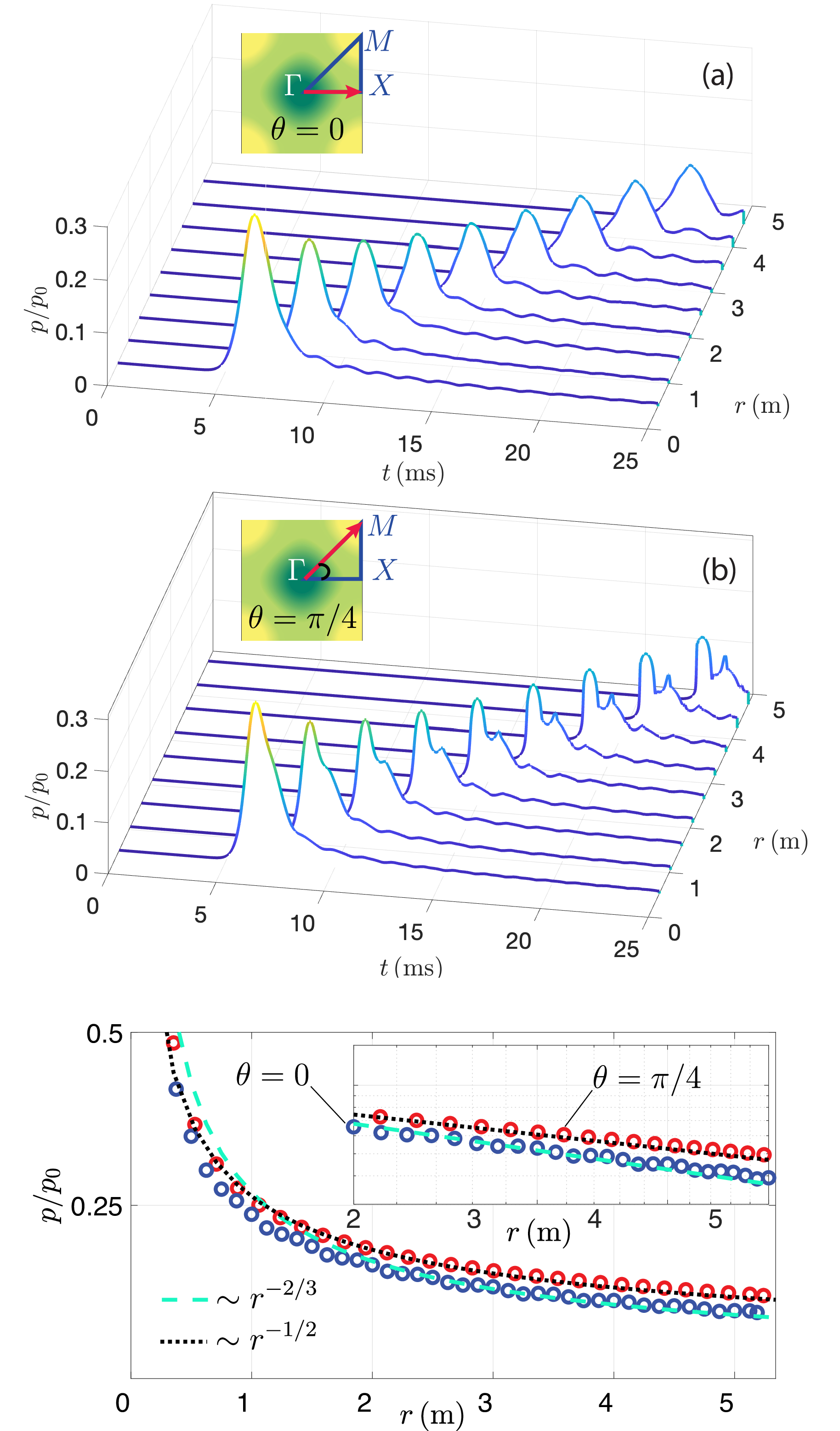} 
 \caption{(a) Evolution of the high-amplitude pulse along the direction $\theta=0$ at fixed distances. (b) Same along the direction $\theta=\pi/4$. (c) Amplitude decay along the direction $\Gamma X$ and $\Gamma M$; numerical results are presented in circles for the direction $\theta=0$ (blue) and $\theta=\pi/4$ (red) and the analytical predictions of Eq.\,\eqref{airy solution square} in dashed (cyan) line, and Eq.\,\eqref{transport square} dotted (black). The inset depicts a zoom of the amplitude decay in log-log scales, which is a straight line of slope $-2/3$ for the direction $\theta=0$ and $-1/2$ for $\theta=\pi/4$. } \label{temporal nonlinear square}
\end{center}
\end{figure}
%

Finally we note that 
the decay rate of a radially symmetric shock wave depends on its dimensionality \cite{landau1945shock,sachdev1986evolution} (see also Ch. 9 in \cite{Whitham} and Ch. 10 in \cite{landaufluid}). For cylindrical shock waves the asymptotic decay rate is $\sim r^{-3/4}$. However, before the shock formation, the amplitude decay varies due to the competition of the effects of curvature and nonlinearity, following the decay law $\sim r^{-1/2}$ 
[see Eq.~\eqref{characteristics}].

\subsubsection{
Numerical results -- nonlinear regime}
To verify the consequent soliton-
and shock-
behavior of the cylindrical wave, we corroborate our theoretical findings 
with numerical simulations. 

To incorporate nonlinear effects, we choose to solve the 2D Westervelt equation
\begin{eqnarray}
&&p_{tt}-c_0^2\Delta p-\frac{\beta}{\varrho c_0^2}\left(p^2\right)_{tt}
\nonumber \\
&=&\nabla\left\{\delta\left[\left(1-\mathrm{k}\right)+\mathrm{k}|\nabla p_t|^{q-1}\right]\nabla p_t\right\},
\label{2d westervelt}
\end{eqnarray}
with $\partial_n p=0$ on the walls. Notice that on the right hand side 
an artificial damping term is introduced, 
in order to cancel higher-order harmonic generation in regions of sharp slopes, as in the case of shock fronts. This particular technique is known as q-Laplacian \cite{nikolic},
where ${\delta}{=2\times10^{-5}\,\text{m}^2/\text{s}}$ is 
the damping coefficient, while $\mathrm{k}\in[0\,1)$ is the fitting parameter, and $\mathrm{q}$ is the order of the nonlinear damping. We note that for our analysis the parameters chosen were $\mathrm{q}=2$ and $\mathrm{k}=2\times10^{-7}$.

The system is 
supplemented with the boundary condition 
\eqref{BC square}, with 
amplitude $p_0=40$\,kPa and 
standard deviation $\sigma=0.6$\,ms, corresponding to a half-width $\approx 1.4$~ms. Contour plots of the resulting high-amplitude field are presented in  Fig.\,\ref{contour nonlinear square} at times $t=[10,\,15,\,20,\,25]$\,ms, in panels~(a-d) respectively. As in the linear case, one can observe that the width and the amplitude of the pulse vary with the direction of propagation, with its maximum amplitude lying in the diagonal $\theta=\pi/4$. The width of the pulse along the same direction is quite narrow 
$\approx 2$ lattice sites), and there is a prominent tail behind it.  On the other hand, for $\theta \neq\pi/4$ the amplitude of the pulse decays faster and the main pulse becomes wider. In particular, its minimum amplitude and maximum duration 
$\approx 4$ lattice sites) occurs for $\theta=0$, as predicted by the analytical solution,  Eq.\,\eqref{soliton approximate square}.

The evolution of the high-amplitude pulse along the directions $\theta=0$ and $\theta=\pi/4$ is also depicted in the 3D plots of Fig.~\ref{temporal nonlinear square}(a) and (b) respectively. For 
$\theta=0$, each of the individual snapshots
at fixed distances 
features a smooth main pulse followed by 
small-amplitude radiation,
and an increasing width due to balance of dispersion, nonlinearity and curvature, predicted by the soliton solution of Eq.\,\eqref{soliton approximate square}. 
In panel (b) the pulse features completely different behavior, as it develops a steep wavefront as predicted by the analytical solution 
eqref{transport square}. In addition, the wave develops a main lobe of temporal half-width corresponding to the unit cell length, $\approx0.73$\,ms, and a slowly decaying tail with second sharp peak. 

The amplitude decay along both directions $\theta=0$ and $\theta=\pi/4$ is presented in
Fig.~\ref{temporal nonlinear square}(c). 
The numerical results are 
depicted by circles for 
$\theta=0$ (blue) and $\theta=\pi/4$ (red), and the analytical predictions by a  dashed (cyan) line 
[Eq.\,\eqref{soliton approximate square}] 
and a dashed (cyan) line 
and a dotted (black) line [Eq.\,\eqref{characteristics}]. The inset 
shows a zoom of the amplitude decay in log-log scales, which is a straight line of slope $-2/3$ for 
$\theta=0$ and $-1/2$ for $\theta=\pi/4$. It is clear that there is a very good agreement between the theoretical predictions and the numerical results.
%
Finally, we note that at the end of the simulation the ratio of the amplitudes between the directions $\theta=0$, and $\theta=\pi/4$ is approximately $\approx 1.3$.


\section{Conclusions}

In this work, we investigated the anisotropic propagation of cylindrical waves in a square network of acoustic waveguides. 
Due to the intrinsic anisotropy of the square lattice band structure, the dispersion relation varies with the propagation direction, giving rise to a rich spectrum of angular-dependent wave phenomena. As a result, the network supports cylindrical waveforms whose character ranges from smooth pulse-shaped solitary waves  
to sharp, shock-like structures, depending on direction.

To describe these effects, we developed a refined analytical framework combining the electroacoustic analogue, 
a supercell formulation, and the transfer matrix method. This way, we derived an 
improved 
2D Boussinesq equation that accurately captures both the dispersive and nonlinear dynamics of the lattice within the monomodal approximation in each waveguide, while explicitly retaining the directional dependence of the dispersion relation --a key ingredient for modeling anisotropic cylindrical waves.

Analytical solutions of the effective model were obtained in both linear and nonlinear regimes. In the low-amplitude limit, lattice anisotropy deforms the Airy-type self-similar cylindrical solution known from isotropic networks into an angle-dependent family of dispersive waves, with 
angle-dependent amplitude decay and spatial spreading. In the limit $\theta \to \pi/4$, the solution smoothly recovers the far-field behavior of the 
2D wave equation, in excellent agreement with direct numerical simulations of the square lattice. In the nonlinear regime, the model predicts anisotropic cylindrical pulses of the form of pulse-shaped solitary waves 
and pulses with shock-front profiles. These predictions were confirmed by numerical simulations of the 
2D Westervelt equation, demonstrating the robust formation and persistence of high-amplitude anisotropic cylindrical waves.

Overall, the results demonstrate that intrinsic lattice anisotropy plays a 
key role in shaping both linear and nonlinear wave evolution in square acoustic networks. The coexistence of soliton-like and shock-like dynamics along different propagation directions highlights the interplay between anisotropic dispersion and nonlinear steepening. More broadly, the modeling framework that was 
developed 
in this work, provides a systematic and predictive pathway for understanding and engineering direction-dependent wave phenomena in multidimensional acoustic metamaterials, with potential extensions to resonant networks, higher-dimensional geometries, and experimental realizations.

\appendix
\section{Electroacoustic analogue}\label{appendix electroacoustic}

\begin{figure}[tb!]
\begin{center}
\includegraphics[width=0.5\textwidth]{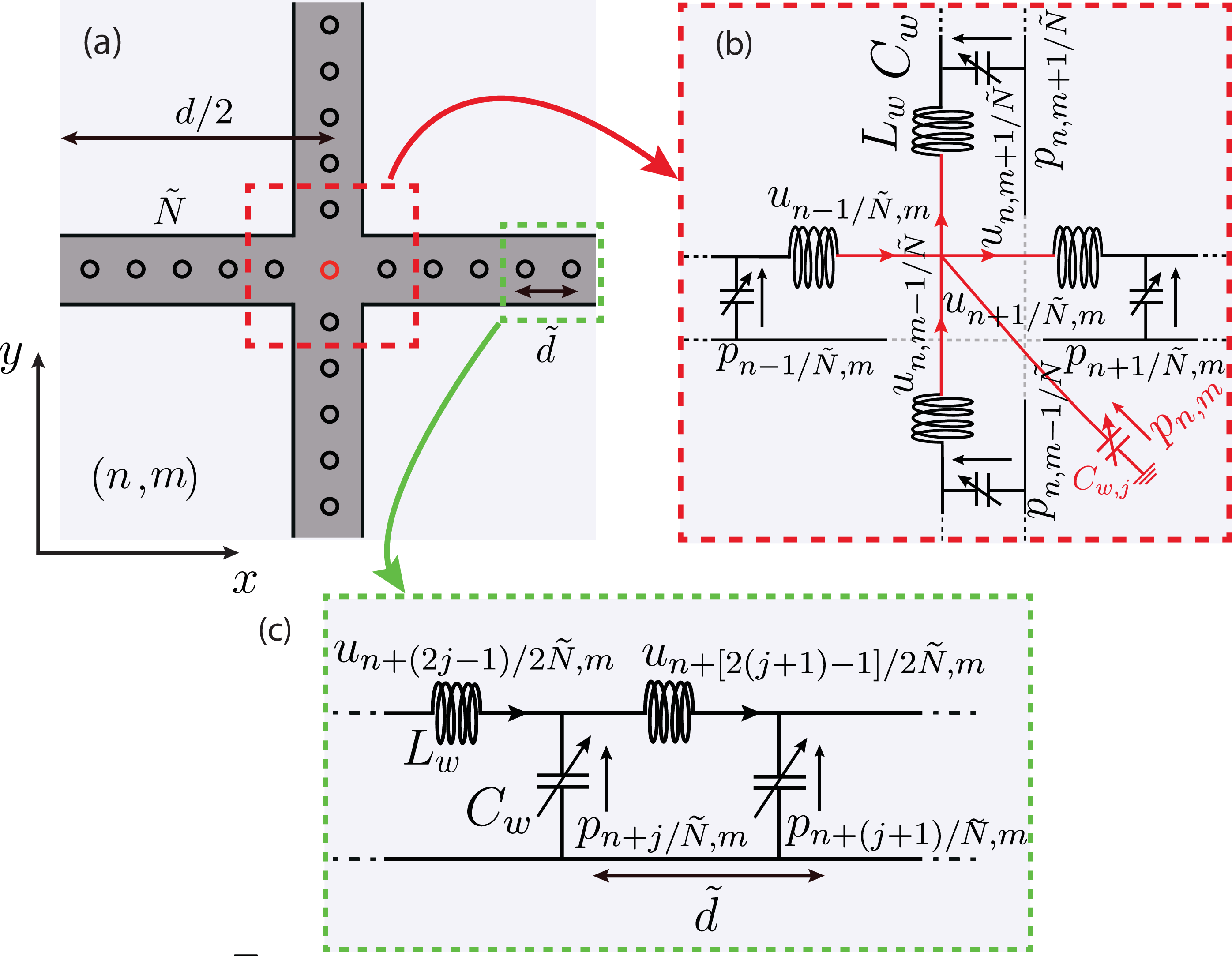}
  \caption{Electroacoustic (EA) analogue. (a) The unit cell of the square network with $\tilde{N}$ discrete points per waveguide segment. (b) The EA representation of the junction of the square network. (c) Same as (b) but for a waveguide segment.   \label{sketch square}}
  \end{center}
\end{figure}

The EA relies on a finite-difference discretization of the simplified conservation laws (\ref{mass simplified waveguide square}–\ref{mass simplified square}). In this framework, we assume that the pressure field varies slowly from junction to junction, an assumption valid only within the long-wavelength approximation. As a result, the discrete nature of the EA, with each junction corresponding to a red circle in Fig.\,\ref{sketch square}(a), introduces additional dispersion into the transmission line equations. Consequently, the accuracy of the EA scheme, which uses one point per unit cell, is of the order $\mathcal{O}\left(d^2\right)$, where $d$ is the lattice distance.

This level of accuracy was acceptable in the case of the network with the Helmholtz resonators (both in 1D and 2D) \cite{vassos_soliton,ioannou2025ring}, or other resonant elements \cite{kinezoula_dark,kinezoula_gap}, since the resonance frequency $\omega_0$ defined the frequency range of interest, ensuring that the EA approximation remained sufficiently precise in the desired range. However, for the square network 
that features an anisotropic dispersion, an accuracy limited to $\mathcal{O}\left(d^2\right)$ is inadequate to capture dispersive effects up to the Bragg frequency $\omega_B = \pi c_0 / d$.  In particular, minimizing the 
discretization-induced 
dispersion is a key point as regards the  
the $\Gamma M$ direction: along this direction, 
dispersion vanishes, which suggests the possibility of emergence of shock waves. 
Therefore, 
artificial dispersion introduced by coarse discretization must be carefully controlled to ensure that 
analytics and numerics reproduce the actual behavior of the network.

To improve accuracy, we adopt a more refined discretization, with $\tilde{N}$ discrete flux points per waveguide segment, as depicted by the black circles in Fig.\,\ref{sketch square}(a), which is similar to that employed in the numerical schemes of \cite{sougleridis2023acoustic,ioannou2025ring}. 
First, we consider the discretization of the unit cell of the square network, as depicted in Fig.\,\ref{sketch square}. In panel (a) we present the unit cell and its discretization, with $\tilde{N}$ discrete points per waveguide segment, for which the conservation laws are considered. The red circle corresponds to the pressure at the junction, while black circles denote a pressure point inside the waveguide segments. Illustrated in panels (b) and (c) are the transmission line representation of the junctions and the waveguide segments, respectively. 
 For simplicity, we shall henceforth refer to the number of discrete flux points as the \textit{order} of the supercell. The error of this enhanced discretization scheme is 
 $\propto \mathcal{O}(d^2/\tilde{N}^2)$; for a sufficient 
 agreement between the dispersion of the effective transmission line and the dispersion relation obtained from TMM Eq.\,\eqref{dispersion square} 
 up to the Bragg frequency, the supercell order should be sufficiently high, i.e., $\tilde{N}\gg1$.
In what follows, we consider the $\tilde{N}_{\text{th}}$ supercell order and derive a general formula that holds for any $\tilde{N}$. 

Kirchhoff current law (KCL), corresponding to mass conservation, for each junction $(n,m)$ yields
\begin{align}
&u_{n, m-1/\tilde{N}}+u_{n-1/\tilde{N}, m}-u_{n+1/\tilde{N}, m}-u_{n, m+1/\tilde{N}}\nonumber\\
&=\frac{d}{d t}\left[C_{w,j}p_{n, m}\right], \label{current N square}
\end{align}
where $u_{n\pm 1/\tilde{N},m\pm 1/\tilde{N}}$ is the acoustic flux per distance $\tilde{d}=d/\tilde{N}$, and $p_{n,m}$ is the pressure at each junction. In addition, we model the capacitance of the junction, $C_{w,j}$ as a pressure-dependent capacitance, which is nonlinear due to the presence of the quadratic term in the equation of state Eq.~\eqref{quadratic}. In particular, we approximate the capacitance of the junction, $C_{w,j}$ as
\begin{equation}
C_{w,j}\approx 2C_{w0}(1-bp_{i,j}),\quad  \nonumber
\end{equation}
where $C_{w0}=\tilde{d}S_w /\varrho_0 c_0^2$ is the linear part of the capacitance and $b=\beta_0/\varrho_0 c^2_{0}$ is the nonlinearity coefficient. 

Next, KCL for each waveguide yields
\begin{align}
  &u_{n, m-[2(\tilde{N}-j)+1]/2\tilde{N}}-u_{n, m-[2(\tilde{N}-j-1)+1]/2\tilde{N}}\nonumber\\
&=\frac{d}{d t}\left[C_{w}p_{n, m-(\tilde{N}-j)/\tilde{N}}\right], \label{current waveguide1 N square}\\
&u_{n, m+(2j-1)/2\tilde{N}}-u_{n, m+[2(j+1)-1]/2\tilde{N}}\nonumber\\
&=\frac{d}{d t}\left[C_{w}p_{n, m+j/\tilde{N}}\right] 
, \label{current waveguide2 N square}\\ 
 &u_{n-[2(\tilde{N}-j)+1]/2\tilde{N}, m}-u_{n-[2(\tilde{N}-j-1)+1]/2\tilde{N}, m}\nonumber\\
&=\frac{d}{d t}\left[C_{w}p_{n-(\tilde{N}-j)/\tilde{N}, m}\right], \label{current waveguide3 N square} \\
 &u_{n+(2j-1)/2\tilde{N}, m}-u_{n+[2(j+1)-1]/2\tilde{N}, m}\nonumber\\
&=\frac{d}{d t}\left[C_{w}p_{n+j/\tilde{N}, m}\right], \label{current waveguide4 N square}  
\end{align}
for $j=\{1,2,...,\tilde{N}-1\}$,
where $C_{w}=C_{w,j}/2$ is the capacitance for a waveguide segment of length $\tilde{d}$.

Furthermore, Kirchhoff voltage law (KVL), corresponding to the momentum conservation, for each waveguide segment reads
\begin{align}
&\!\!\!\!\!\!\!\!\!\!\!\!\!\!\!\!\!\!p_{n,m-(\tilde{N}+1-j)/\tilde{N}}-p_{n,m-(\tilde{N}-j)/\tilde{N}}\nonumber\\
&=L_w \frac{d}{d t} u_{n, m-[2(\tilde{N}-j)+1]/2\tilde{N}},\label{voltage waveguide1 N square}\\
&\!\!\!\!\!\!\!\!\!\!\!\!\!\!\!\!\!\!p_{n,m+j/\tilde{N}}-p_{n,m+(j+1)/\tilde{N}}\nonumber\\
&=L_w \frac{d}{d t} u_{n, m+(2j-1)/2\tilde{N}},\label{voltage waveguide2 N square}\\
&\!\!\!\!\!\!\!\!\!\!\!\!\!\!\!\!\!\!p_{n-(\tilde{N}+1-j)/\tilde{N},m}-p_{n-(\tilde{N}-j)/\tilde{N},m}\nonumber\\
&=L_w \frac{d}{d t} u_{n-[2(\tilde{N}-j)+1]/2\tilde{N},m}, \label{voltage waveguide3 N square}\\
&\!\!\!\!\!\!\!\!\!\!\!\!\!\!\!\!\!\!p_{n+(\tilde{N}+1+j)/\tilde{N},m}-p_{n+(\tilde{N}+j+1)/N,m}\nonumber\\
&=L_w \frac{d}{d t} u_{n+(2j-1)/2\tilde{N},m} .\label{voltage waveguide4 N square} 
\end{align}
Note that 
KVL can only be 
considered in the waveguide segments, since the junction is approximated as a single node. 

Combining the Kirchhoff laws for the junctions, Eqs.\,(\ref{current N square})-(\ref{voltage waveguide4 N square}), and 
keeping leading-order dispersive and nonlinear terms, we obtain the following differential difference equation (DDE) for the pressure \eqref{DDE 4th square}. As mentioned in the main text, $\alpha_{\tilde{N}}$ is the dispersion coefficient associated with the 4th-order time derivative, with a distinct value at each supercell order $\tilde{N}$. Here we present different values of the supercell dispersion coefficient $\alpha_{\tilde{N}}$
\begin{align}
    \alpha_2=\frac{1}{16},\,\, \alpha_3=\frac{2}{27},\,\, \alpha_4=\frac{5}{64},\,\,\alpha_5=\frac{2}{25},\,\, \alpha_{10}=\frac{33}{400}. \nonumber
\end{align}
The relative error $\delta_N=|\alpha_N-\alpha_{TM}|/\alpha_{TM}$ between the coefficients of the long-wavelength approximation of the transfer matrix \eqref{square longwavelength} and the continuum approximation of each supercell transmission line is
\begin{align}
\delta_2=0.25,\,\, \delta_3=0.11,\,\, \delta_4=0.063,\,\, \delta_5=0.04,\,\, \delta_{10}=0.01\nonumber.
\end{align}

\section{Derivatives in polar coordinates}\label{appendix derivatives}
The Laplacian, biharmonic and 4th order spatial derivative operators in polar coordinates are
\begin{align}
    &\Delta= \partial_{rr}+ \frac{1}{r}\partial_r+\frac{1}{r^2}\partial_{\theta\theta},\nonumber\\
 &\Delta^2=\partial_{r r r r}+\frac{2}{r^2} \partial_{r r \theta \theta}+\frac{1}{r^4} \partial_{\theta \theta \theta \theta}+\frac{2}{r} \partial_{r r r}-\frac{2}{r^3} \partial_{r \theta \theta}\nonumber\\
 &-\frac{1}{r^2} \partial_{r r}+\frac{4}{r^4} \partial_{\theta \theta}+\frac{1}{r^3} \partial_r\nonumber.
    \end{align}
\begin{widetext}
\begin{align}
&\partial_{xxxx}+\partial_{yyyy}=\frac{3+\cos{(4\theta)}}{4}\partial_{rrrr}-\frac{\sin{(4\theta)}}{r}\partial_{rrr\theta}+\frac{3\left[1-\cos{(4\theta)}\right]}{2r^2}\partial_{rr\theta\theta}
+\frac{\sin{(4\theta)}}{r^3}\partial_{r\theta\theta\theta}+\frac{3+\cos{(4\theta)}}{4r^4}\partial_{\theta\theta\theta\theta}\nonumber\\
    &\!\!\!+\frac{3\left[1-\cos{(4\theta)}\right]}{2r}\partial_{rrr}+\frac{6\sin{(4\theta)}}{r^2}\partial_{rr\theta}-\frac{3\left[1-5\cos{(4\theta)}\right]}{2r^3}\partial_{r\theta\theta}-\frac{3\sin{(4\theta)}}{r^4}\partial_{\theta\theta\theta}-\frac{3\left[1-5\cos{(4\theta)}\right]}{4r^2}\partial_{rr}-\frac{14\sin{(4\theta)}}{r^3}\partial_{r\theta}\nonumber\\
    &+\frac{3-11\cos{(4\theta)}}{r^4}\partial_{\theta\theta}+\frac{3\left[1-5\cos{(4\theta)}\right]}{4r^3}\partial_{r}+\frac{12\sin{(4\theta})}{r^4}\partial_{\theta}. \nonumber
\end{align}
\end{widetext}

\section{Optimal dispersion coefficients}
\label{appendix optimal dispersion}
The generalized improved 2D Boussinesq Eq.\,\eqref{Boussinesq improved square} derived in Section III features the undetermined coefficients $\alpha_x$, $\alpha_x^\prime$, $\alpha_t$ and $\alpha_m$. A procedure for determining these coefficients has been proposed in \cite{wautier} for 2D periodic media, as part of a 2nd-order homogenization technique of a 2D chessboard medium, while the same approach has been applied to 1D periodic layered media in \cite{cornaggia}.

According to \cite{wautier,cornaggia}, it is possible to determine the \textit{optimal} dispersion coefficients, for which the improved Boussinesq Eq.\, \eqref{Boussinesq improved square} does not encounter the spurious dispersion problem of the ill-posed Boussinesq Eq.\,\eqref{Boussinesq square} while simultaneously approximating the exact dispersion relation of the TMM \eqref{dispersion square}. 
 
First, the dispersion relation \eqref{dispersion improved square} can further be simplified in the long-wavelength approximation, i.e., for  $\omega\ll\omega_B$; indeed, upon Taylor expanding Eq.\,\eqref{dispersion improved square}, we obtain (for outgoing waves)
 \begin{align}   
    &\omega\frac{d}{c_0}\approx\frac{c}{c_0}(kd)+\left(\frac{c}{c_0}\right)^3\alpha(\theta)\left(kd\right)^3+\left(\frac{c}{c_0}\right)^5\alpha(\theta)
    \nonumber\\
&\times\frac{12\alpha_m+28\alpha_t+4\alpha_x+\left[3+\cos{(4\theta)}\right]\alpha_x^\prime
    }{16}\left(kd\right)^5,
    \label{taylor boussineq square}
 \end{align}
with
\begin{align}
\alpha(\theta)=\frac{4\left(\alpha_t+\alpha_m-\alpha_x\right)-\left[3+\cos{(4\theta)}\right]\alpha^\prime_x}{8}.
\end{align}
On the other hand, in the long-wavelength approximation, the dispersion relation of the square lattice can be approximated--by Taylor expanding \eqref{dispersion square}
\begin{align}
\omega(qd) \approx&\frac{c_0}{\sqrt{2}d}\Big\{q d-\frac{1}{96}\left[1+\cos (4 \theta)\right](q d)^3+\frac{1+\cos{(4\theta)}}{96}\nonumber\\
&\times\frac{\left[17+\cos{(4\theta)}\right]}{192}(qd)^5\Big\} \label{Taylor square2}.
\end{align}

By letting the dimensionless Bloch wavenumber $qd$, to be equivalent to the dimensionless wavenumber $kd$, and matching the asymptotic expressions of the improved Boussinesq dispersion relation \eqref{taylor boussineq square}, and the TMM \eqref{Taylor square2}, we obtain the following constraints for the dispersion coefficients 
\begin{align}
&\alpha_t+\alpha_m-\alpha_x=\frac{1}{12}, \quad \alpha_{x}^\prime=\frac{1}{6},
&\alpha_x+3\alpha_m+7\alpha_t=\frac{7}{12} \label{coefficients square}.
\end{align}
Notice, that the anisotropic dispersion coefficient is already determined.
In addition, as we did for the 1D periodic waveguide, we consider that the group velocity at the edge of the Brillouin zone, along the direction $\theta=0$, vanishes 
\begin{align}
\frac{\partial\omega(kd,0)}{\partial (kd)}\Big|_{kd=\pi}=0. \label{group velocity}
\end{align}
Hence, the number of constraints \eqref{coefficients square} and \eqref{group velocity} is equal to the number of unknowns. In the case where $\alpha_x,\,\alpha_m,\,\alpha_t\neq0$, the system is solved numerically. 
However, although a great agreement is observed between the two models \cite{wautier}, the onset of the Bangap  for the improved Boussinesq equation dispersion is not as accurate as expected. To overcome this difficulty, we consider $\alpha_t=0$ and then apply the constraints
\begin{align}
&\alpha_t+\alpha_m-\alpha_x=\frac{1}{12}, \quad \alpha_{x}^\prime=\frac{1}{6},\quad\frac{\partial\omega(kd,0)}{\partial (kd)}\Big|_{kd=\pi}=0 \label{coefficients square2},
\end{align}
where the system of equations is solved analytically
\begin{eqnarray} \alpha_x&=&\frac{48-3\pi^2-\pi\sqrt{96+\pi^2}}{24\pi^2},
\\ \alpha_m&=&\frac{48-\pi^2-\pi\sqrt{96+\pi^2}}{24\pi^2}.
\end{eqnarray}
The corresponding dispersion relation is plotted in dotted (green) line. The latter is clearly a better approximation of the TMM dispersion relation, which solidifies the validity of Eq.\,\eqref{Boussinesq improved square2} in the long-wavelength and low-frequency regime. A comparison of the improved Boussinesq for $\alpha_t=0$, the TMM, and the continuum approximation of the supercell is also shown in Fig.\,\ref{boussinesq dispersion}.

\bibliography{bibliographie.bib}

\end{document}